\documentclass[11pt]{article}

\usepackage{graphicx}
\usepackage{amsmath}
\usepackage{amsfonts}
\usepackage{amssymb}
\allowdisplaybreaks
\usepackage{physics}
\usepackage{microtype}
\usepackage{bbold}
\usepackage{cite}
\usepackage{tabu}
\usepackage{multirow}
\usepackage{hhline}
\usepackage{arydshln}
\usepackage{braket}

\usepackage{xcolor}
\usepackage{framed}
\definecolor{shadecolor}{rgb}{0.9,0.9,0.9}
\usepackage{etoolbox}

\usepackage[normalem]{ulem}

\usepackage[colorlinks=true, linkcolor=black, filecolor=blue, urlcolor=black, citecolor=blue]{hyperref}

\newcommand{\clo}{\mathcal{O}}
\newcommand{\clf}{\mathcal{F}}
\newcommand{\bbk}{\mathbb{K}}
\newcommand{\bbf}{\mathbb{F}}
\newcommand{\bbs}{\mathbb{S}}

\newcommand{\bbi}{\mathbb{1}}
\newcommand{\nn}{N_C}
\newcommand{\gb}{\Gamma_b}
\newcommand{\eqspace}[1]{\phantom{.}\!\hspace{#1}}

 \newcommand{\bea}{\begin{eqnarray}}
\newcommand{\eea}{\end{eqnarray}}
\newcommand{\be}{\begin{equation}}
\newcommand{\ee}{\end{equation}}
\newcommand{\ba}{\begin{align}}
\newcommand{\ea}{\end{align}}

\newcommand{\M}{\mathcal{M}}

\newcommand\rref[1]{(\ref{#1})}

\newlength{\slength}
\newcommand{\zw}{Z^{\textnormal{wormhole}}_{g=2\times g=2}}
\newcommand{\zg}{Z^{\textnormal{grav}}_{g=2\times g=2}}
\newcommand{\EH}{S_{\textnormal{EH}}}

\newcommand{\sbmatrix}[1]{
{\tiny\arraycolsep=0.1\arraycolsep\ensuremath{\begin{bmatrix}#1\end{bmatrix}}}
}

\newcommand{\fker}[6]{
\mathbb{F}_{#1 #2}{\tiny\arraycolsep=0.2\arraycolsep\ensuremath{\begin{bmatrix}#4 & #3 \\ #5 & #6\end{bmatrix}}}
}

\newcommand{\fbloc}[4]{
\clf {\tiny\arraycolsep=0.2\arraycolsep\ensuremath{\begin{matrix}#2#1 \\ #3#4\end{matrix}}}\:
}

\newcommand{\dbloc}[4]{
\rho {\tiny\arraycolsep=0.2\arraycolsep\ensuremath{\begin{matrix}#2#1\\#3#4\end{matrix}}}\:
}

\newcommand{\sker}[3]{
\mathbb{S}_{#1 #2}[#3]
}

\numberwithin{equation}{section}

\begin{document}

\begin{flushright}
\hfill{\tt CERN-TH-2021-166}
\end{flushright}

\vskip10mm

\begin{center} {\LARGE   Non-Gaussianities in the Statistical Distribution \\ \vspace{0.3em} of Heavy OPE Coefficients and Wormholes}

\vskip10mm

Alexandre Belin$^{1}$, Jan de Boer$^{2}$, Diego Liska$^{2}$  
\vskip1em
{\it 1) CERN, Theory Division, 1 Esplanade des Particules, Geneva 23, CH-1211, Switzerland} \\ \vspace{0.5em}
{\it 2)  Institute for Theoretical Physics and $\Delta$-Institute for Theoretical Physics, University of Amsterdam, PO Box 94485, 1090GL Amsterdam, The Netherlands} \\
\vskip5mm

\vskip5mm

\tt{a.belin@cern.ch, J.deBoer@uva.nl, d.liska@uva.nl  }

\end{center}

\vskip10mm

\begin{abstract}

The Eigenstate Thermalization Hypothesis makes a prediction for the statistical distribution of matrix elements of simple operators in energy eigenstates of chaotic quantum systems. As a leading approximation, off-diagonal matrix elements are described by Gaussian random variables but higher-point correlation functions enforce non-Gaussian corrections which are further exponentially suppressed in the entropy. In this paper, we investigate non-Gaussian corrections to the statistical distribution of heavy-heavy-heavy OPE coefficients in chaotic two-dimensional  conformal field theories. Using the Virasoro crossing kernels, we provide asymptotic formulas involving arbitrary numbers of OPE coefficients from modular invariance on genus-$g$ surfaces. We find that the non-Gaussianities are further exponentially suppressed in the entropy, much like the ETH. We discuss the implication of these results for products of CFT partition functions in gravity and Euclidean wormholes. Our results suggest that there are new connected wormhole geometries that dominate over the genus-two wormhole.

\end{abstract}

\newpage
\phantom{a}
\vspace{-4em}

\tableofcontents

\newpage

\section{Introduction}

The Eigenstate Thermalization Hypothesis (ETH) \cite{PhysRevA.43.2046,PhysRevE.50.888} is expected to be one of the most universal properties of chaotic quantum many-body systems, which captures important aspects of thermalization (see \cite{DAlessio:2016rwt} for a review). The ETH ansatz stipulates that the matrix elements of simple operators (like few-qubit operators on chaotic spin chains) in energy eigenstates adopt a universal structure
\be
\label{ETHanzats}
\bra{E_i}O\ket{E_j} = f_O(\bar{E}) \delta_{ij} + g_O(\bar{E},\omega) e^{-S(\bar{E})/2} R_{ij} \,,
\ee
where $\bar{E}$ and $\omega$ are the mean energy and energy differences, while $f_O$ and $g_O$ are smooth functions and $S$ is the microcanonical entropy. $R_{ij}$ are erratic pseudo-random numbers which are fixed for any given Hamiltonian, but are often taken to be independent random variables with a Gaussian distribution of zero mean and unit variance. This ansatz guarantees that thermal (or microcanonical) one and two-point functions are correctly reproduced. While an actual proof of the ETH remains to be found, it can be checked numerically (see for example \cite{PhysRevE.82.031130,Sonner:2017hxc}) and is widely expected to hold.

In recent years, it has been understood that taking the variables $R_{ij}$ to be uncorrelated is in fact only an approximation, and can lead to inconsistencies if taken too seriously. For example, this Gaussian ansatz with independent random variables will fail to reproduce higher-point functions and, in particular, out-of-time-ordered correlation functions \cite{Foini:2018sdb,PhysRevLett.122.220601,PhysRevLett.123.230606} (see also \cite{Dymarsky:2018ccu,Richter:2020bkf} for other issues that follow from having independent Gaussian variables). Assuming connected four-point functions to be of order unity\footnote{In holographic CFTs, the fact that connected correlation functions are $1/N^2$ suppressed can lead to extra polynomial suppressions in the entropy for the higher moments.}, \cite{Foini:2018sdb} constrained the higher moments of the matrix elements $O_{ij}$ with certain cyclic contractions
\be
\overline{O_{i_1i_2}O_{i_2i_3} ..... O_{i_{k}i_1}} \sim e^{-(k-1)S} \label{FK1} \,,
\ee
which gives a net weight for the connected $k$-th moment of $O$ to be of order
\be
O_{ij}^{\text{k-th moment}} \sim e^{-\frac{k-1}{k}S}  \label{FK2}\,.
\ee
In this sense, one can view the ETH ansatz with Gaussian random variables as a leading approximation that gets corrected by higher moments further suppressed in the entropy. It is important to emphasize that even though the higher moments are suppressed, they can lead to competing effects in correlation functions due to the extra sums over intermediate states.

Applying ETH-type ideas to chaotic conformal field theories is of great interest for several reasons. It ties in with the understanding of quantum chaos in continuum quantum field theory and has far-reaching consequences for quantum gravity through the AdS/CFT correspondence \cite{Maldacena:1997re}. Indeed, thermalization of a holographic CFT is closely connected to  black hole formation in the bulk. The state/operator correspondence of CFTs gives a particularly nice interpretation to the ETH ansatz \cite{Lashkari:2016vgj}\footnote{A subtlety of the ETH in CFTs is that due to the presence of conformal symmetry, one should distinguish primary and descendant operators. It is easy to see that descendants violate the ETH ansatz, so one should really view it as applying to primary states only. This question becomes particularly subtle for 2d CFTs due to Virasoro symmetry \cite{Maloney:2018yrz,Dymarsky:2018lhf,Dymarsky:2019etq,Datta:2019jeo,Besken:2019bsu}, but we still expect ETH to hold when applied to Virasoro primaries.}
\be
\bra{E_i}O^a\ket{E_j}= \bra{0}O_i O^a O_j\ket{0}\equiv C_{ij}^a \,,
\ee
since matrix elements of local operators in energy eigenstates are given by OPE coefficients, where the scaling dimensions of the operators creating the state $\Delta_{i,j}\to\infty$ (this is the thermodynamic limit). On the other hand, the scaling dimension $\Delta_a$ of the simple operator is held fixed. The ETH ansatz is therefore a statement about the statistical distribution of OPE coefficients. 

However, conformal field theories raise questions that do not have a natural counterpart in quantum mechanics. It is interesting to ask what the statistical distribution of other OPE coefficients is, when either one or all three of the operators are taken to be heavy. We call these OPE coefficients $C_{abi}$ and $C_{ijk}$. One may be tempted to view these as the expectation value of a very complicated operator in a simple low-energy eigenstate and as the expectation value of a complicated operator in a high energy state, respectively. Unfortunately, thermalization does not obviously offer any insight into the structure of these statistical distributions. Nevertheless, we expect that in a chaotic theory high energy states should be very difficult to distinguish. This led to the conjecture that all OPE coefficients involving heavy operators also have a pseudo-random distribution which is described to leading order by a Gaussian \cite{Belin:2020hea}.\footnote{This conjecture is meant to apply to chaotic CFTs, by which we mean any CFT that displays level repulsion in the spectrum of primary operators (this is equivalent to displaying a linear ramp in the spectral form factor). In $d=2$, level repulsion should be understood for Virasoro primaries only. Level repulsion may be equivalent to the CFT having $c>1$ and no extended chiral algebra, but there
currently is no proof of such a statement.} This conjecture is particularly relevant for holographic CFTs, for which correlations of operator matrix elements and OPE coefficients have attracted much interest due to the connection to wormholes \cite{Saad:2019pqd,Pollack:2020gfa, Belin:2020hea,Blommaert:2020seb,Belin:2020jxr,Altland:2021rqn,Freivogel:2021ivu,Goto:2021mbt}. 

Proving ETH or the OPE randomness hypothesis (ORH) is of course extremely difficult. Unlike quantum mechanics where one can get numerical evidence for ETH by direct diagonalization, this is extremely difficult for conformal field theories which always have an infinite-dimensional Hilbert space. A direct numerical test of ETH using the conformal bootstrap is still far beyond reach.\footnote{Some progress has been achieved analytically in large $c$ 2d CFTs, assuming the dominance of the Virasoro identity block in heavy states (see for example \cite{Fitzpatrick:2014vua}). Note that this probes the diagonal entries of ETH, but does not say anything about the individual off-diagonal matrix elements.} However, crossing symmetry (and modular invariance in $d=2$) highly constrains the distribution of OPE coefficients and one can extract averaged OPE and spectral densities from these constraints. These averaged OPE and spectral densities are known as asymptotic formulas \cite{Pappadopulo:2012jk,Kraus:2016nwo,Das:2017vej,Cardy:2017qhl,Das:2017cnv,Qiao:2017xif,Mukhametzhanov:2018zja,Pal:2019zzr,Brehm:2018ipf,Romero-Bermudez:2018dim,Hikida:2018khg,Collier:2019weq}, the most famous of which is the Cardy formula \cite{cardyformula}. It is important to emphasize that these formulas are in no way proofs of the ETH, since they cannot describe individual OPE coefficients and, in fact, they apply to all CFTs including free or integrable theories. However, they are consistency checks to show that on average, one can obtain distributions \textit{compatible} with the ETH, and therefore offer important insights in the absence of numerics. The status of known asymptotic formulas for the Gaussian part of the distribution of OPE coefficients is reported in Table \ref{tab:status}.

\begin{table}[t]
\begin{center}
\begin{tabular}{lllll} 
\hline
                           & \multicolumn{2}{l}{$d=2$}                               & \multicolumn{2}{l}{$d>2$}                                \\ 
\hline
$C_{LLH}$                  & $\overline{|C_{LLH}|^2}$  & $\checkmark\phantom{\Big(}$ & $\overline{|C_{LLH}|^2}$  & $\checkmark\phantom{\Big(}$  \\ 
\hdashline[1pt/1pt]
\multirow{2}{*}{$C_{LHH}$} & $\overline{C_{LHH}}$      & $\checkmark\phantom{\Big(}$ & $\overline{C_{LHH}}$      & $\checkmark\phantom{\Big(}$  \\
                           & $\overline{|C_{LHH'}|^2}$ & $\checkmark\phantom{\Big(}$ & $\overline{|C_{LHH'}|^2}$ & $\checkmark\phantom{\Big(}$  \\ 
\hdashline[1pt/1pt]
$C_{HHH}$                  & $\overline{|C_{HHH}|^2}$  & $\checkmark\phantom{\Big(}$ & \multicolumn{2}{c}{?}                                    \\
\hline
\end{tabular}
% \begin{tabular}{c|c|c}
%   &   $d=2$ &   $d>2$ \\[5pt]
% \hline
% \\[-0.5em]
% $C_{LLH}$ & $\overline{|C_{LLH}|^2} \ \ \ \checkmark$ & $\overline{|C_{LLH}|^2} \ \ \ \checkmark$  \\[5pt] \hline \\[-0.5em]
% \multirow{2}{*}{$C_{LHH}$} & $\overline{C_{LHH}} \ \ \ \checkmark$ & $\overline{C_{LHH}} \ \ \ \checkmark$ \\[5pt]
%     & $\overline{|C_{LHH'}|^2} \ \ \ \checkmark$ & $\overline{|C_{LHH'}|^2}$ \\[5pt]
% \hline \\[-0.5em]
% $C_{HHH}$ & $\overline{|C_{HHH}|^2} \ \ \ \checkmark$ & ? \\[5pt]
% \hline
% \end{tabular}
\end{center}
\caption{The status of the various asymptotic formulas known in CFT for light (L) or heavy (H) operators. It only remains to constrain $C_{HHH}$ in $d>2$, which will be discussed in \cite{6ptasymptotics}.}
\label{tab:status}
\end{table}

In this paper, we will address the following question: how do the higher moments of the distribution of OPE coefficients behave, and how do asymptotic formulas constrain such distributions? To the best of our knowledge, this is currently unexplored. We will focus on CFTs in $d=2$ and study higher moments of OPE coefficients with three heavy operators, i.e. the $C_{ijk}$. We will establish formulas for these OPE coefficients similar to \rref{FK1} and \rref{FK2} which were relevant for ETH. The observables we use to constrain these distributions are higher genus partition functions. At the technical level, we rely on the Virasoro crossing kernel \cite{Ponsot:1999uf,Ponsot:2000mt}, extending the results obtained for the Gaussian part of the distribution in \cite{Collier:2019weq}. For holographic CFTs, we will then investigate the implications of these higher moments for the gravitational dual, in particular for wormhole geometries.

\subsection{Summary of results}

\begin{figure}
    \centering
    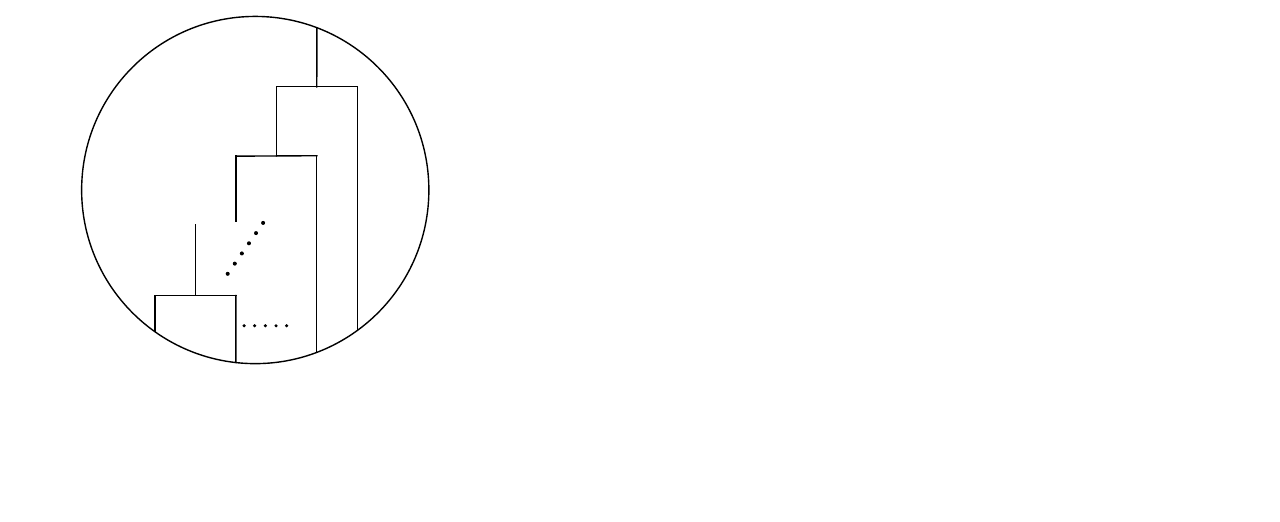
    \caption{On the left the trivalent graph associated with the configuration of OPE coefficients in the skyline channel. On the right, the comb channel.}
    \label{comb and skyline}
\end{figure}

We study the higher moments of two families of OPE coefficients in two-dimensional CFTs. The configurations we consider originate from two different conformal block decompositions of the genus-$g$ partition function. The two different decompositions correspond to the ``skyline" and ``comb" channels depicted in Figure  \ref{comb and skyline}. We will focus on irrational, compact, and unitary CFTs -- i.e. theories without an extended chiral algebra that have $c>1$, a discrete spectrum and an infinite number of primary states -- with a unique $\mathfrak{sl}(2)$-invariant ground state.

Our first result is a formula for the average value of OPE coefficients in the skyline channel\footnote{Throughout this paper, the symbol $\approx$ means that in the limit of interest, the two expressions agree up to a multiplicative constant. }
\begin{equation}
\label{firstresult}
  \overline{C_{\{\textnormal{skyline}\}}}  \approx \left(\frac{27}{16}\right)^{3\frac{\nn}{2}\Delta_1}e^{\left(2-5\frac{\nn}{2}\right)\pi\sqrt{\frac{c-1}{3}\Delta_1}}\Delta_1^{\frac{5c-11}{72}\nn}, \qquad \Delta_i\gg c, J_i, |\Delta_i-\Delta_j|,
\end{equation}
where $\nn$ is the number of OPE coefficients in the average and $J_i = |h_i-\bar h_i|$ is the spin of the primary operators $\clo_i$. Note that $\nn$ is always an even number. This formula is valid for $\nn\geq2$; however, it is interesting to note that it reduces to Cardy's formula when $\nn = 0$. This result is the higher genus generalization of the formulas derived in \cite{Collier:2018exn} from the genus-two partition function. Our result is universal in the sense that it only depends on the central charge of the CFT and not on any other details of the theory. In section \ref{section 3.1}, we use the modular invariance of the genus-$g$ partition function to show that this equation follows from the dominance of the identity operator in the appropriate cross-channel. 

Our second result is a similar formula for the average value of OPE coefficients, this time in the comb channel
\begin{equation}
\label{secondresult}
    \overline{C_{\{\textnormal{comb}\}}} \approx \left(\frac{27}{16}\right)^{3\frac{\nn}{2}\Delta_1}e^{\left[\frac{3}{2}\left(1-3 \frac{\nn}{2}\right)\pi \sqrt{\frac{c-1}{3}}-\sqrt{2}\left(\frac{\nn}{2}+1\right)\pi (\alpha_\chi + \bar \alpha_\chi )\right]\sqrt{\Delta_1}}\Delta_1^{\frac{5c-11}{72}\nn}, \quad \Delta_i\gg c, J_i, |\Delta_i-\Delta_j|,
\end{equation}
where $\alpha_\chi$ is a Liouville variable, defined in section \ref{section 2}, and $\chi$ is the lightest primary operator in the theory that is not the identity. This time, the result is not a universal theory-independent formula, but there is an explicit dependence on the light data of the theory. As we will show, this comes from the fact that the contribution of the identity vanishes in the cross-channel and we must thus go to the first non-trivial primary. A similar observation was made in \cite{Kraus:2016nwo} for the average of $C_{aij}$ from torus one-point functions. A difference with the torus one-point function is that no OPE coefficient of $\chi$ appears in the asymptotic formula (other than $C_{1\chi \chi}$ which is one in our conventions). This means that it is only important that $\chi$ exists, but it is unimportant how it couples to other operators.

Throughout this paper, we limit our discussion to the case where the differences between the weights of all heavy operators are held fixed. However, the techniques we use to extract these results also work when considering other asymptotic limits. For instance, in the skyline channel,  \rref{Eq:recursionSkyline} only assumes that the weights are large, i.e. $h_i\gg c$. Thus, we could also use this formula to study, for example, the limit where the conformal weight ratios are held fixed instead of their differences. Moreover, we have cited the results here as a function of scaling dimension, but the master formula \rref{Eq:recursionSkyline} is a function of the left and right moving weights. Therefore, one could also use it to explore the large-spin limit.

These asymptotic formulas should be understood as valid once averaged over a suitably large energy band and so far, we have not stated what range of primary states we must average over. A precise answer to this question can be given using Tauberian theorems; see \cite{Qiao:2017xif,Mukhametzhanov:2018zja,Pal:2019zzr,Mukhametzhanov:2019pzy,Mukhametzhanov:2020swe,Das:2020uax}  for recent applications of Tauberian theorems in this context. While the asymptotic formulas are universal, the minimal window over which one should average can be strongly theory-dependent, and the interpretation of the asymptotic formulas depend on whether the theory is chaotic or not. For example, the minimal window will be much smaller for chaotic theories than integrable ones. If we make the extra assumption that the CFT is chaotic, we expect the typical OPE coefficients to be rather close to the averaged value and we thus hope that the asymptotic formula gives a good estimate for the individual coefficients.

We would also like to emphasize that the technology we develop in this paper works for more general observables. For example, using the sequence of transformations we describe in Figures \ref{recursion skyline process} and \ref{recursion comb process}, we can easily find statistics for a broader class of families. The configuration that is shown in Figure \ref{other recursion} is one such example. It would be interesting to keep exploring the dynamics of two-dimensional CFTs at large scaling dimension using these methods, and eventually give a complete characterization of asymptotic formulas for 2d CFTs. We leave this for future work.

This paper is organized as follows. In section \ref{section 2} we present the notation that we will use throughout this paper and review the Virasoro fusion and modular kernels with the example of the four-point function on the sphere and the torus partition function. We discuss the Moore-Seiberg construction of crossing kernels for general observables and use it to derive the OPE statistics coming from the genus-three partition function. The main results of this paper are derived in section \ref{section 3}, where we apply this technology to study the higher moments of the OPE configurations associated with the skyline channel and the comb channel. We also discuss the extent to which we can apply our methods to obtain a broader set of statistics. In section 4, we investigate the consequences of these non-Gaussianities for the square of partition functions and discuss the wormhole interpretation of our results. We finish with some conclusions in section \ref{sec:discussion}. We present most of the technical details about the elementary fusion and modular kernels in the appendices.

\section{Crossing kernels and crossing equations}
\label{section 2}

This section introduces the Virasoro fusion and modular kernels coming from the sphere four-point function and the torus one-point function, respectively. We use these examples to set the notation and highlight the general strategy we use to find OPE statistics. 

In what follows, we will write the central charge $c$ in terms of the \emph{background charge} $Q$ or the \emph{Liouville coupling} $b$ defined by $c = 1 + 6 Q^2$ and $Q = b + b^{-1}$. For $c>25$, we choose to work with $0<b<1$. To label the representations, besides $h$, we also use $P$ and $\alpha$:
\begin{equation}
    h = \frac{Q^2}{4} + P^2 = \alpha(Q-\alpha), \quad
    \alpha = \frac{Q}{2} + i P. 
\end{equation}
These definitions are redundant since they are invariant under the exchange of $P \rightarrow -P$ or $\alpha \rightarrow Q- \alpha$. We will be working with the following conventions. For $h\leq (c-1)/24$, that is $h\leq Q^2/4$, we choose $\alpha \in [0,Q/2]$ and $P \in i [0,Q/2]$. In particular, $h=0$ corresponds to $\alpha = 0$ and $P =  i Q/2$. We will call this regime the \emph{discrete} range. For  $h\geq (c-1)/24$ we choose $\alpha \in Q/2 + i \mathbb{R}$ and real $P$; we call this the \emph{continuum} range. We will refer to the variables $\alpha$ and $P$ as \emph{momenta}. Whenever possible, we will use $\alpha$ in the discrete range and $P$ in the continuum. The use of the Liouville variables will simplify the expressions in the rest of the paper.

\subsection{The fusion kernel}
\label{section 2.1}
\begin{figure}
    \centering
    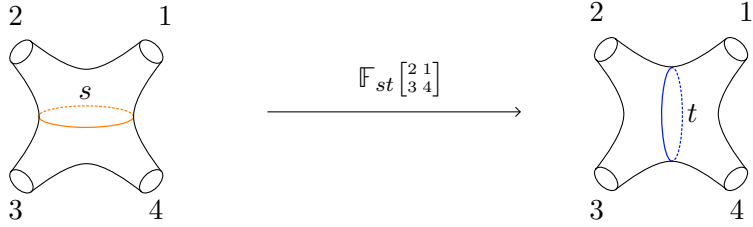
    \caption{The two conformal block decompositions of the four-point function with four different primary operators. The fusion kernel is the \emph{change-of-basis} matrix relating the two bases of conformal blocks in the s- and t- channel.}
    \label{fourpoint}
\end{figure}

To start, we revisit the arguments given in \cite{Collier:2019weq,Collier:2018exn} for the four-point function. We will work in a slightly more general setting considering the four-point function depicted in  Figure \ref{fourpoint} with four, potentially different, primary external operators. For now, we focus on the case where the four external operators are in the continuum range: $h_i>(c-1)/24$. In the next section, we will study what happens when some of the external operators are in the discrete range. 

Recall that we can decompose the four-point function in terms of s- or t-channel conformal blocks
\begin{equation}
\begin{split}
    \langle \clo_1(0) \clo_2(z,\bar z)\clo_3(1)\clo_4'(\infty)\rangle &= \sum_{\clo_s} C_{12s}C_{34s}
    \fbloc{1}{2}{3}{4}(P_{\clo_s};z)
    \fbloc{\bar 1}{\bar 2}{\bar 3}{\bar 4}(\bar P_{\clo_s};\bar z)
    \\ &= \sum_{\clo_t} C_{14t}C_{23t}
    \fbloc{3}{2}{1}{4}(P_{\clo_t};1-z)
    \fbloc{\bar 3}{\bar 2}{\bar 1}{\bar 4}(\bar P_{\clo_t};1-\bar z),
\end{split}
\end{equation}
where the sum runs over all the primary operators in the theory. We define the spectral density $\dbloc{1}{2}{3}{4}(P_s,\bar P_s)$ of OPE coefficients as the distribution satisfying 
\begin{equation}
    \int_{-\infty}^\infty \frac{dP_s}{2}\frac{d\bar P_s}{2}  \dbloc{1}{2}{3}{4}(P_s,\bar P_s) \clf(P_s) \bar \clf(P_s) = \sum_{\clo_s} C_{12s}C_{34s}\clf(P_{\clo_s}) \bar \clf(\bar P_{\clo_s}).
\end{equation}
Here $\clf(P_s)$ does not refer to a specific conformal block, but it refers to any function that can be written as a linear combination of the s- or t-channel conformal blocks. Note the distinction between the continuous variable $P_s$ and the discrete one $P_{\clo_s}$. The integration is defined over the real axis, and the density can be viewed as a sum of delta functions with support at real or imaginary values of $P$\footnote{For a discussion about delta functions with imaginary support, we recommend the appendices of \cite{Maxfield:2019hdt}.}. To account for the symmetry $P \rightarrow -P$ we have included a factor of one half in the measure. 

The fusion kernel (also known as the F-transform, crossing kernel or $6j$ symbol) $\bbf_{st}$ relating the two conformal block decompositions is defined as the Fourier-like transform satisfying
\begin{equation}
\label{defFusionKernel}
    \fbloc{3}{2}{1}{4}(P_t;1- z) = \int_{-\infty}^\infty \frac{dP_s}{2} \fker{P_s}{P_t}{P_1}{P_2}{P_3}{P_4} \fbloc{1}{2}{3}{4}(P_s;z).
\end{equation}
We can think of the conformal blocks as a basis of a vector space that contains the four-point correlation function. The s- and t-channel conformal blocks simply correspond to two different bases of the same vector space. The Virasoro fusion kernel is then defined as the \emph{change-of-basis} matrix relating the two decompositions. The fusion kernel is formally defined when all external operators have momenta in the continuum, but we will analytically continue away from this regime to study arbitrary primary operators. The remarkable fact is that this kernel was written down explicitly in \cite{Ponsot:2000mt,Ponsot:1999uf} by Ponsot and Teschner. We study its exact structure, properties, and different limits in appendix \ref{appendixfusion}. 

The crossing equation is derived by plugging in equation \rref{defFusionKernel}  into the integral form of the four-point function; we find that
\begin{equation}
    \label{CrossEqFourPoint}
\begin{split}
    \dbloc{1}{2}{3}{4}(P_s,\bar P_s) &= \int_{-\infty}^\infty \frac{dP_t}{2}\frac{d\bar P_t}{2} 
    \fker{P_s}{P_t}{P_1}{P_2}{P_3}{P_4}
    \fker{\bar P_s}{\bar P_t}{\bar P_1}{\bar P_2}{\bar P_3}{\bar P_4}  \dbloc{3}{2}{1}{4}(P_t,\bar P_t)\\
    &=\sum_{\clo_t}  C_{14t}C_{23t}
    \fker{P_s}{P_{\clo_t}}{P_1}{P_2}{P_3}{P_4}
    \fker{\bar P_s}{\bar P_{\clo_t}}{\bar P_1}{\bar P_2}{\bar P_3}{\bar P_4}.
\end{split}
\end{equation}
By itself, formula \rref{CrossEqFourPoint} does not make much sense. This is because the sum does not converge in the usual sense. Since the density of OPE coefficients is a sum of delta functions, it does not have a smooth asymptotic behavior. This sum only converges in the sense of distributions and requires some smearing against an appropriate function. We will write many asymptotic formulas in this paper for the OPE densities, but these should always be understood as being smeared over some energy window. This approach is familiar when discussing the density of states, i.e. Cardy's formula. Formally, the density of states is not a function, it is a sum of delta functions and thus a distribution. The smooth formula we consider when we write Cardy's formula is a smeared version of the true density of states. For more details, see \cite{Mukhametzhanov:2019pzy}. 

An important property of the crossing equation is that the first non-vanishing term in the sum dominates in the heavy limit $P_s,\bar P_s \rightarrow \infty$. The technical result required to show this is
\begin{equation}
\label{asymptoticfusionkernel}
\log \fker{P_s}{P_t}{1}{2}{3}{4}  = -2 P_s^2 \log 4 + \pi(Q-2\alpha_t) P_s + 2\left[-\frac{c+1}{8}+\sum_{i=1}^4 h_i\right]\log P_s + \clo(1), \quad P_s \rightarrow \infty.
\end{equation}
This equation was derived in \cite{Collier:2018exn} for the case of identified external operators $\clo_1 = \clo_4$ and $\clo_2 = \clo_3$. The generalization for arbitrary external operators is straightforward and discussed in appendix \ref{appendixfusion}. Equation \rref{asymptoticfusionkernel} implies that, in the heavy limit, the sum in the formula \rref{CrossEqFourPoint} is well approximated by
\begin{equation}
\label{OPEfourpoint}
    \dbloc{1}{2}{3}{4}(P_s,\bar P_s) \approx C_{14\chi}C_{23\chi}
    \fker{P_s}{P_\chi}{P_1}{P_2}{P_3}{P_4}
    \fker{\bar P_s}{\bar P_\chi}{\bar P_1}{\bar P_2}{\bar P_3}{\bar P_4}.
\end{equation}
Here, $\chi$ is the lightest operator in the theory that couples to the external operators $\clo_i$ in the sense that $C_{14\chi}C_{23\chi}\neq 0$. Corrections to this equation are exponentially suppressed in $P_s$.

To extract the average value of the OPE coefficients $\overline{C_{12s} C_{34s}}$ from equation \rref{OPEfourpoint}, we note the following relationship between densities 
\begin{equation}
\int \frac{dP_s}{2}\frac{d\bar P_s}{2}\; \dbloc{1}{2}{3}{4}(P_s,\bar P_s)\clf(P_s)\bar \clf(\bar P_s) =:\int \frac{dP_s}{2}\frac{d\bar P_s}{2}\; \rho_0(P_s)\rho_0(\bar P_s)\; \overline{C_{12s}C_{34s}}\clf(P_s)\bar \clf(\bar P_s),
\end{equation}
where  $\rho_0$ corresponds to the density of primary operators. At large $P_s$, we know from Cardy's formula that 
\begin{equation}
    S(P_s) = \log \rho_0(P_s) = 2\pi Q P_s+\clo(1).
\end{equation}
Hence, after stripping off the density of states from formula \rref{OPEfourpoint}, we have that
\begin{equation}
\label{fourPointStatistics}
    \overline{C_{12s} C_{34s}} \approx C_{14\chi}C_{23\chi} 16^{-\Delta_s}e^{-\pi(Q+2\alpha_\chi)P_s}P_s^{2\left(-\frac{c+1}{8}+\sum_{i=1}^4 h_i\right)}e^{-\pi(Q+2\bar \alpha_\chi)\bar P_s}{\bar P_s}^{2\left(-\frac{c+1}{8}+\sum_{i=1}^4 \bar h_i\right)}.
\end{equation}
An interesting case is when the external operators are identified in pairs: $\clo_1=\clo_4$ and $\clo_2 = \clo_3$. This allows us to set $\alpha_\chi = \alpha_\bbi = 0$ and leads to a universal formula for the average value of $\overline{C_{12s}^2}$, see \cite{Collier:2018exn,Collier:2019weq}. If the operators are not identical, we have a non-universal formula for the average value a slightly more general configuration of OPE coefficients. Corrections to these equations are exponentially suppressed in $P_s$. Equation \rref{fourPointStatistics} is only valid when $\chi$ is in the discrete range. If $\alpha_\chi \in Q/2 + i \mathbb{R}$, then one can no longer argue using equation \rref{asymptoticfusionkernel} that the first non-vanishing term in the formula \rref{CrossEqFourPoint} dominates the sum, as the other terms are no longer exponentially suppressed.\footnote{This result can be seen from equation \rref{asymptoticfusionkernel}, where the exponential suppression comes from the linear piece $-2\pi\alpha_t P_s$. The lightest operator will thus dominate in the sum, as long as $\Re(\alpha_t)< Q/2$. However, if the lightest operator already has $\Re(\alpha_t) = Q/2$, then we no longer have an exponential suppression and cannot separate the contribution of the lightest operator from all the other ones.}.

In the following sections, we will use the same arguments to prove more general statistics. The computations will involve different limits of the fusion kernel. It is important to keep in mind that this kernel has a rich structure, and its behavior changes depending on the situation we are considering. We derive and review all the technical results we need about this kernel in the appendices. 

\subsubsection{Discrete range external operators}
When the external operators $\clo_i$ are sufficiently light, $h_i < (c-1)/24$, the fusion kernel develops a new subtlety arising from the poles in the internal momentum $P_s$. For generic external dimensions and operators, the fusion kernel has eight semi-infinite lines of poles extending in the upper half-plane and eight other semi-infinite lines extending in the lower half-plane. These poles happen when $P_s = \pm i [Q/2+i(P_1+P_2) + nb +mb^{-1}]$, $P_s = \pm i [Q/2+i(P_3+P_4) + nb +mb^{-1}]$, for $n,m\in \mathbb{Z}_{\geq 0}$, or at different permutations of these zeros under the reflections $P_i \rightarrow -P_i$. When $P_1$, $P_2$ or $P_3$, $P_4$ are light enough, e.g. $\Re(Q/2 + i[P_1+P_2]) < 0$ and $\Im(P_1),\Im(P_2)>0$, the poles of the kernel cross the integration contour. To maintain analyticity in the parameters, we must deform the contour and include portions surrounding the relevant poles, see \cite{Collier:2018exn} for more details. These poles contribute to the integral in equation \rref{defFusionKernel}:
\begin{equation}
    \fbloc{3}{2}{1}{4}(P_t;1-z)
     = - 2\pi \sum_m \underset{P_s = P_m}{\Res}\{\bbf_{P_sP_t}   \fbloc{1}{2}{3}{4}(P_s;z)\} +  \int_{-\infty}^\infty \frac{dP_s}{2} \fker{P_s}{P_t}{1}{2}{3}{4} \fbloc{1}{2}{3}{4}(P_s;z),
\end{equation}
where the sum runs over all the poles at imaginary $P_m$ crossing the contour of integration. 

The details of these new contributions are not relevant for our analysis. For this paper, it suffices to note that we can add these terms explicitly in the crossing kernel \cite{Maxfield:2019hdt}:
\begin{equation}
    \fker{P_s}{P_t}{1}{2}{3}{4} \rightarrow  \fker{P_s}{P_t}{1}{2}{3}{4} + \sum_{m,a} x^a_m \delta^{(a)}(P_s-P_m).
\end{equation}
Here, $x_m^a =x_m^a(P_i;P_t)$  are functions independent of $P_s$ and we are fixing the contour of integration to be over the real axis i.e., without contour deformations. For example, if the poles are simple, only $x_m^1 = \Res_{P_s=P_m}\{\bbf_{st}\}$ is nonzero; for double poles, we have two contributions: $x_m^1 =\Res_{P_s=P_m}\{\bbf_{st}\}$ and $x_m^2 =\lim_{P\rightarrow P_m} (P-P_m)^2\bbf_{Pt}$. Similar expressions can be derived for higher poles. Unless there is a fine-tuning on the external parameters, such as $i(P_2 + P_1) = -Q/2-nb-mb^{-1}$, we do not expect to find poles at real values of $P_m$. Moreover, poles are confined in the region $P_3+P_4 \leq P_s \leq P_1+P_2$. Hence, in the heavy limit, $P_s\gg P_i$ we can ignore these contributions. For the results we derived in the previous section, this implies that \rref{fourPointStatistics} also holds for external operators in the discrete regime.

\subsection{The modular kernel}
\label{section 2.2}
\begin{figure}
    \centering
    %% Creator: Inkscape 1.0 (4035a4fb49, 2020-05-01), www.inkscape.org
%% PDF/EPS/PS + LaTeX output extension by Johan Engelen, 2010
%% Accompanies image file 'torusonepointfunction.pdf' (pdf, eps, ps)
%%
%% To include the image in your LaTeX document, write
%%   \input{<filename>.pdf_tex}
%%  instead of
%%   \includegraphics{<filename>.pdf}
%% To scale the image, write
%%   \def\svgwidth{<desired width>}
%%   \input{<filename>.pdf_tex}
%%  instead of
%%   \includegraphics[width=<desired width>]{<filename>.pdf}
%%
%% Images with a different path to the parent latex file can
%% be accessed with the `import' package (which may need to be
%% installed) using
%%   \usepackage{import}
%% in the preamble, and then including the image with
%%   \import{<path to file>}{<filename>.pdf_tex}
%% Alternatively, one can specify
%%   \graphicspath{{<path to file>/}}
%% 
%% For more information, please see info/svg-inkscape on CTAN:
%%   http://tug.ctan.org/tex-archive/info/svg-inkscape
%%
\begingroup%
  \makeatletter%
  \providecommand\color[2][]{%
    \errmessage{(Inkscape) Color is used for the text in Inkscape, but the package 'color.sty' is not loaded}%
    \renewcommand\color[2][]{}%
  }%
  \providecommand\transparent[1]{%
    \errmessage{(Inkscape) Transparency is used (non-zero) for the text in Inkscape, but the package 'transparent.sty' is not loaded}%
    \renewcommand\transparent[1]{}%
  }%
  \providecommand\rotatebox[2]{#2}%
  \newcommand*\fsize{\dimexpr\f@size pt\relax}%
  \newcommand*\lineheight[1]{\fontsize{\fsize}{#1\fsize}\selectfont}%
  \ifx\svgwidth\undefined%
    \setlength{\unitlength}{300.47244094bp}%
    \ifx\svgscale\undefined%
      \relax%
    \else%
      \setlength{\unitlength}{\unitlength * \real{\svgscale}}%
    \fi%
  \else%
    \setlength{\unitlength}{\svgwidth}%
  \fi%
  \global\let\svgwidth\undefined%
  \global\let\svgscale\undefined%
  \makeatother%
  \begin{picture}(1,0.24131293)%
    \lineheight{1}%
    \setlength\tabcolsep{0pt}%
    \put(0,0){\includegraphics[width=\unitlength,page=1]{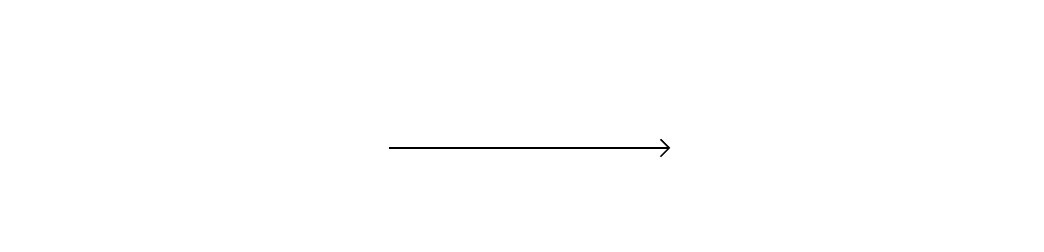}}%
    \put(0.18888896,0.21787767){\color[rgb]{0,0,0}\makebox(0,0)[lt]{\lineheight{1.25}\smash{\begin{tabular}[t]{l}$\clo$\end{tabular}}}}%
    \put(0.9309448,0.21787767){\color[rgb]{0,0,0}\makebox(0,0)[lt]{\lineheight{1.25}\smash{\begin{tabular}[t]{l}$\clo'$\end{tabular}}}}%
    \put(0.00614322,0.21787767){\color[rgb]{0,0,0}\makebox(0,0)[lt]{\lineheight{1.25}\smash{\begin{tabular}[t]{l}$0$\end{tabular}}}}%
    \put(0.69235548,0.21787767){\color[rgb]{0,0,0}\makebox(0,0)[lt]{\lineheight{1.25}\smash{\begin{tabular}[t]{l}$0$\end{tabular}}}}%
    \put(0.47007971,0.13109936){\color[rgb]{0,0,0}\makebox(0,0)[lt]{\lineheight{1.25}\smash{\begin{tabular}[t]{l}$\sker{P}{P'}{0}$\end{tabular}}}}%
    \put(0,0){\includegraphics[width=\unitlength,page=2]{torusonepointfunction.pdf}}%
  \end{picture}%
\endgroup%

    \caption{The modular kernel relating the two different conformal block decompositions of the torus one-point function.}
    \label{torus1point}
\end{figure}

We now introduce the Virasoro modular kernel or S-transform and briefly review how to derive Cardy's formula from it, see \cite{McGough:2013gka,Benjamin:2019stq,Kusuki:2019gjs,Collier:2018exn,Collier:2019weq,Maxfield:2019hdt,Mukhametzhanov:2019pzy} for more details and applications. 

The modular kernel $\bbs_{PP'}$ arises from the modular invariance of the torus one-point function:
\begin{equation}
\begin{split}
    \langle \clo_0 \rangle_{\tau} &= \sum_{\clo} C_{\clo \clo \clo_0} \clf(P_{\clo};\tau)\bar \clf(\bar P_{\clo}; \bar \tau)\\
    &= \tau^{-h_0} \bar \tau^{-\bar h_0}\sum_{\clo'} C_{\clo'\clo' \clo_0} \clf(P_{\clo'};-1/{\tau})\bar \clf(\bar P_{\clo'}; -1/{\bar\tau}),
\end{split}
\end{equation}
where $\tau$ is the modular parameter of the torus. The two conformal block decompositions are depicted in Figure \ref{torus1point}. The kernel is defined as the Fourier-like transform that decomposes the torus one-point blocks of one frame into its modular transformed frame:
\begin{equation}
    \clf(P';-1/\tau) = \tau^{h_0} \int_{-\infty}^{\infty} \frac{dP}{2} \sker{P}{P'}{P_0} \clf(P;\tau).
\end{equation}
This kernel is known in closed form \cite{Teschner:2003at}; we study its properties in appendix \ref{appendixmodular}. As with the four-point function, from the definition of the modular kernel we can write a crossing equation between OPE densities
\begin{equation}
\label{Eq:crosseqmodular}
\begin{split}
    \rho_{m}(P,\bar P) &= \int\frac{dP'}{2}\frac{d\bar P'}{2} \sker{P}{P'}{P_0} \sker{\bar P}{\bar P'}{\bar P_0} \rho_m(P',\bar P') \\&= \sum_{\clo'} C_{\clo'\clo'\clo_0}
    \sker{P}{P'}{P_0}
    \sker{\bar P}{\bar P'}{\bar P_0}.
\end{split}
\end{equation}
In the heavy limit, $P_s \rightarrow \infty$, this sum is dominated by its first non-vanishing term 
\begin{equation}
\label{Eq:ua15}
    \rho_m(P,\bar P) \approx C_{\chi \chi \clo_0} 
    \sker{P}{\chi}{P_0}\sker{\bar P}{\bar \chi}{\bar P_0}.
\end{equation}
Here,  $\chi$ is the first operator that couples with $\clo_0$ in the sense of having a non-vanishing OPE coefficient. The key result that establishes this formula was derived in \cite{Collier:2019weq} and reads
\begin{equation}
    \log \sker{P}{P'}{P_0} = 2\pi (Q-2\alpha') P + h_0 \log P+\clo(1), \qquad P\rightarrow\infty.
\end{equation}
From equation \rref{Eq:ua15}, we find the following formula for the average value of the torus one-point OPE coefficients
\begin{equation}
\label{eq:torusOPE}
    \overline{C_{\clo\clo \clo_0}} \approx C_{\clo_0 \chi \chi} \frac{\sker{P}{P_\chi}{P_0}\sker{\bar P}{\bar P_\chi}{\bar P_0}}{\rho_0(P)\rho_0(\bar P)}.
\end{equation}

If we set the external operator $\clo_0$ to the identity, the first contribution comes from $\chi = \bbi$. Since $C_{\clo\clo \bbi}=1$, we have that 
\begin{equation}
\label{cardysformula}
\rho_0(P)\rho_0(\bar P) \approx \sker{P}{\bbi}{\bbi}\sker{\bar P}{\bbi}{\bbi} \sim e^{2\pi Q(P+\bar P)}.
\end{equation}
This is, of course, Cardy's formula for the asymptotic density of primary states at large $h$ and $\bar h$. If $\clo_0$ is not the identity, equation \rref{eq:torusOPE} reads 
\begin{equation}
\label{torusonepoint}
    \overline{C_{\clo\clo \clo_0}} \sim C_{\clo_0 \chi \chi} \frac{\sker{P}{P_\chi}{P_0}\sker{\bar P}{\bar P_\chi}{\bar P_0}}{\rho_0(P)\rho_0(\bar P)} \sim  C_{\clo_0 \chi \chi} e^{-4\pi (\alpha_\chi P + \bar \alpha_\chi\bar P)}P^{h_0}\bar P^{\bar h_0}.
\end{equation}
This result was derived in \cite{Kraus:2016nwo,Kraus:2017ezw} using scaling and global blocks and it was then re-derived in \cite{Collier:2019weq} using the modular kernel technology. Corrections to the formulas \rref{cardysformula} and \rref{torusonepoint} are exponentially suppressed at large $P$.  Note that equation \rref{torusonepoint} is only valid if $\chi$, the lightest operator that couples to $\clo_0$, is in the discrete regime where $\alpha$ is real. If $\chi$ is in the continuum, then $\alpha_\chi \in Q/2 + i\mathbb{R}$ and corrections due to the propagation of other operators become relevant.

\subsection{Prelude: the genus-three partition function}
\label{section 2.3}

\begin{figure}
    \centering
    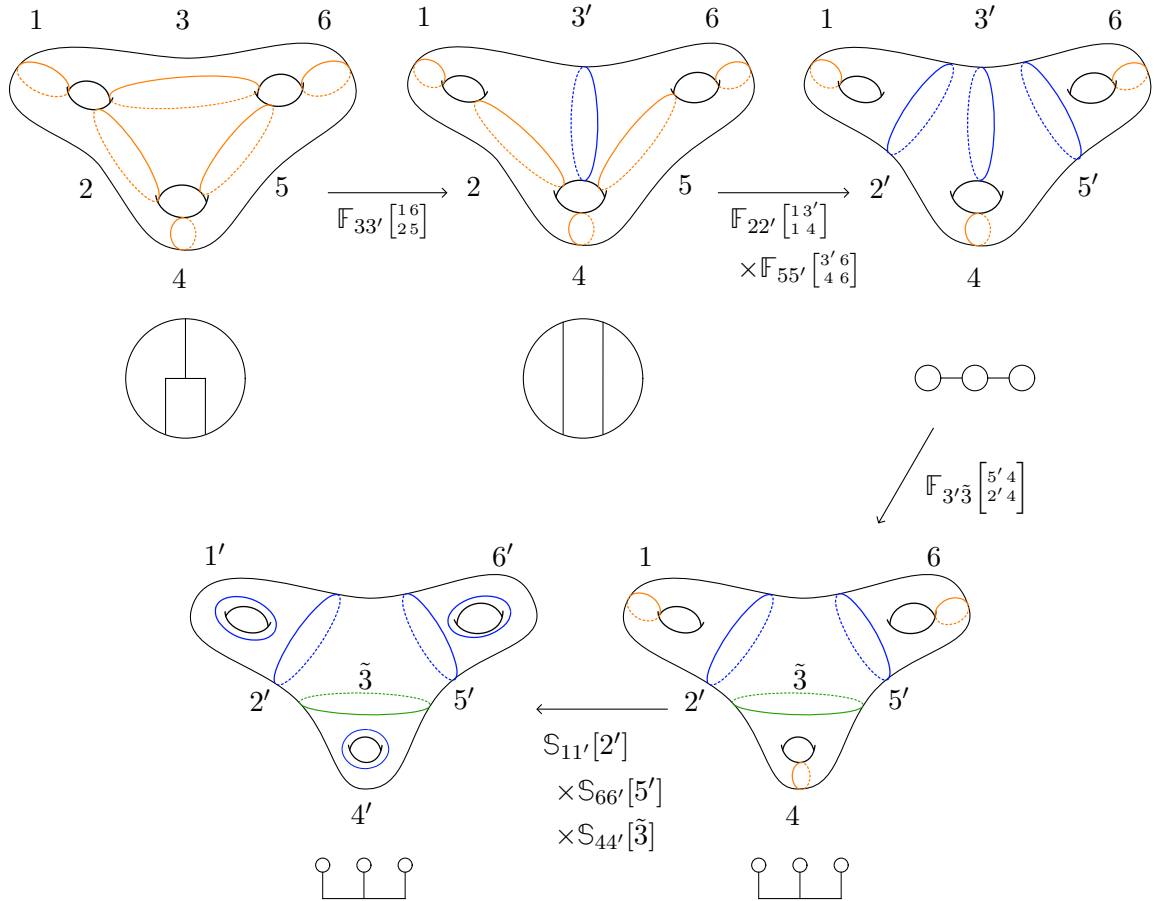
    \caption{The crossing kernels relating the different pair-of-pants decompositions or channels of the genus-three Riemann surface. From left to right the names of these channels are: skyline, sunset, necklace, and comb. The trivalent graphs associated to each decomposition are shown below each surface. Note that the fusion kernel may change the topology of the graph while the modular kernel always leaves the diagram invariant.}
    \label{genus3}
\end{figure}

To discuss more general observables on different Riemann surfaces, we can use the state-operator correspondence to sew together multiple copies of the punctured sphere\footnote{Recall that the three-point coefficients $C_{123}$ correspond to the correlation function $\langle \clo_1\clo_2\clo_3\rangle$ of three operators inserted on the sphere.}. For example, the four-point function is built by sewing two three-point functions at a point, and the torus one-point by sewing two different points on the same sphere. In general, any $n$-point correlation function on a genus-$g$ Riemann surface can be decomposed into a set of $2g+n-2$ three-point functions (topologically equivalent via the operator-state correspondence to a pair of pants). Different cuttings of the same surface correspond to different conformal blocks decompositions; however, all cuttings evaluate to the same correlation function. The general idea is to view the various decompositions as a distinct choice of basis. The crossing kernel responsible for the change of basis can always be built from the elementary fusion and modular kernels using the Moore-Seiberg construction \cite{Moore:1988qv,Moore:1988uz}. We show how this works with an example. 

We consider the genus-three partition function depicted in Figure \ref{genus3}. Each step shows a different cutting of the surface and the elementary crossing kernel relating them. From last to first, we have two \emph{combs}, a \emph{necklace}  and a \emph{sunset} channel decomposition. These are common names found in the literature. We call the initial channel \emph{skyline}. All these channels have their analogs at higher genus. At genus two, the sunset and skyline channel degenerate into the same pair-of-pants decomposition; this also happens with the comb and necklace channels. At genus two, the necklace and comb channels are commonly referred to as the \emph{dumbbell} channel. We will adhere to this terminology throughout this paper.  

The partition function on the skyline and comb channels reads
\begin{equation}
\begin{split}
        Z_3 &= \sum_{\clo_1\dots\; \clo_6} C_{124}C_{136}C_{456}C_{235} \clf_S(P_i) \bar \clf_S(\bar P_i)\\
        &= \sum_{\clo_1'\!\dots\; \tilde \clo_3 \dots \; \clo_6'} C_{1'1'2'}C_{4'4'\tilde 3} C_{6'6'5'}C_{2'\tilde 3 5'}\clf_C(P'_i) \clf_C(\bar P'_i).
\end{split}
\end{equation}
Here, we have suppressed the dependence of the blocks on the moduli since we do not need them. The crossing equation now reads 
\begin{equation}
  \rho_S^{(3)}(P_i) = \int \frac{dP_3'}{2}\int \frac{d\bar P_3'}{2} \sum_{\clo_1'\!\dots\; \tilde \clo_3 \dots \; \clo_6'}  C_{1'1'2'}C_{4'4'\tilde 3} C_{6'6'5'}C_{2'\tilde 3 5'} \bbk_{\{P_i\}\{P_3';P_1'\!\dots\; \tilde P_3 \dots \; P_6'\}} \bbk_{\{\bar P_i\}\{\bar P_3';\bar P_1'\!\dots\; \tilde{\bar P}_3 \dots \; \bar P_6'\}}.
\end{equation}
Where $\rho_S^{(3)}(P_i)$ is the spectral density of OPE coefficients in the sunset channel at genus three. Unlike the previous examples, this time we have an extra step in the transformation, so we are left with a combination of sums and integrals. The full crossing kernel reads
\begin{multline}
    \bbk_{\{P_i\}\{P_3';P_1'\!\dots\; \tilde P_3 \dots \; P_6'\}} =\\ 
    \fker{P_3}{P_3'}{P_6}{P_1}{P_2}{P_5} 
    \fker{P_2}{P_2'}{P_3'}{P_1}{P_1}{P_4}
    \fker{P_5}{P_5'}{P_6}{P_3'}{P_4}{P_6}
    \fker{P_3'}{\tilde P_3}{P_4}{P_5'}{P_2'}{P_4}
    \sker{P_1}{P_1'}{P_2'}
    \sker{P_6}{P_6'}{P_5'}
    \sker{P_4}{P_4'}{\tilde P_3}.
\end{multline}

Note that the three modular kernels exponentially suppress the momenta $1',6'$ and $4'$, which pushes them to be the identity. The variables $2',5'$ and $\tilde 3$ are then also forced to be the identity, but this time by the OPE coefficients $C_{1'1'2'}$, $C_{4'4'\tilde 3}$, and $C_{6'6'5'}$. We see that setting $\clo_1'$, $\clo_4'$ and $\clo_6'$ to the identity automatically sets the operators $\clo_2',\clo_5'$ and $\tilde \clo_{3}$ to the identity as well.

Having all the primed variables but $3'$ set to the identity reveals a delta function in the crossing kernel (see formula \rref{skylinedelta} in the appendix)
\begin{equation}
    \fker{3'}{\bbi}{4}{\bbi}{\bbi}{4} = \delta(P_3' - P_4) e^{\clo(1)}.
\end{equation}
This delta function allows us to evaluate the integral in the crossing equation and shows that the spectral density of OPE coefficients in the skyline channel at genus three is well approximated by 
\begin{equation}
    \rho_{S}^{(3)}(P_i) \sim
    \rho_0(P_1)\rho_0(P_4)\rho_0(P_6)\fker{P_3}{P_4}{P_6}{P_1}{P_2}{P_5} 
    \fker{P_2}{\bbi}{P_4}{P_1}{P_1}{P_4}
    \fker{P_5}{\bbi}{P_6}{P_4}{P_4}{P_6}\times \textnormal{(a.c.)}.
\end{equation}
where a.c. stands for the antiholomorphic counterpart of this equation. This result is valid when $P_1,P_6,P_4,\bar P_1,\bar P_6,\bar P_4\gg c$ and corrections are exponentially suppressed in at least one of these variables.  Stripping off the corresponding densities of OPE coefficients yields the following universal result 
\begin{equation}
\label{Eq:generalOPEgen3}
    \overline{C_{124}C_{136}C_{456}C_{235}} \sim \frac{
    \fker{3}{4}{6}{1}{2}{5}
    \fker{2}{\bbi}{4}{1}{1}{4}
    \fker{5}{\bbi}{6}{4}{4}{6}}
    {{\rho_0(P_2)\rho_0(P_3)\rho_0(P_5)}}\times \textnormal{(a.c.)}.
\end{equation}
Since we have multiple internal weights, we can study this equation in distinct heavy limits. In this paper, however, we are interested in the limit  $P_i = P + \delta_i$ and $P \gg c, \delta_i$. Here, we have the following results for the fusion kernel:
\begin{align}
    \log \fker{3}{4}{6}{1}{2}{5} &= 2(\delta_3-\delta_4)P \log \frac{27}{16} + \clo(1),\\
    \log \fker{2}{\bbi}{4}{1}{1}{4} &= 3 P^2 \log\frac{27}{16}+\left(-\pi Q
    +2(\delta_2+\delta_1+\delta_4)\log\frac{27}{16}\right) P +\frac{5Q^2-1}{6}\log P+\clo(1).
\end{align}
The first equation is derived in appendix \ref{allheavymomenta}, while the second is one of the main results of \cite{Collier:2019weq}. From these expressions, we can write
\begin{equation}
\label{OPEgen3density}
    \log \rho_{S}^{(3)} \sim 6 P^2\log \frac{27}{16}+4\pi Q P + \frac{5Q^2-1}{3}\log P + \textnormal{ (a.c.) },
\end{equation}
and thus conclude that 
\begin{equation}
    \overline{C_{124}C_{136}C_{456}C_{235}}  \approx \left(\frac{27}{16}\right)^{6\Delta}e^{-8\pi Q (P+\bar P)}(P\bar P)^{\frac{5Q^2-1}{3}}.
\end{equation}
Here, we have set $\sum\delta_i = 0$. This result is universal, meaning that it only depends on the central charge of the theory. Note that the exponential suppression corresponds to a factor of $e^{-4S}$, which is further exponentially suppressed compared to the disconnected product of two heavy-heavy-heavy $\overline{C_{123}^2}$ factors, which together would scale as to $e^{-3S}$. This is the first clear sign that the non-Gaussianities are exponentially suppressed in the distribution of OPE coefficients.

\section{Statistics at higher genus}
\label{section 3}

We now study the statistics of OPE coefficients associated with the skyline and comb channels. For this, we use two transformations that simplify these channels at arbitrary genus. At the end of this section, we explain how to use this technology to find the statistics of more general OPE configurations. The observable we use to derive these results is the genus-$g$ partition function. We focus on the heavy limit with large conformal weights and fixed differences -- i.e. $P_i-\delta_i= P$ and $P\gg c,\delta_i$.

\subsection{The skyline channel}
\label{section 3.1}

\begin{figure}
\centering
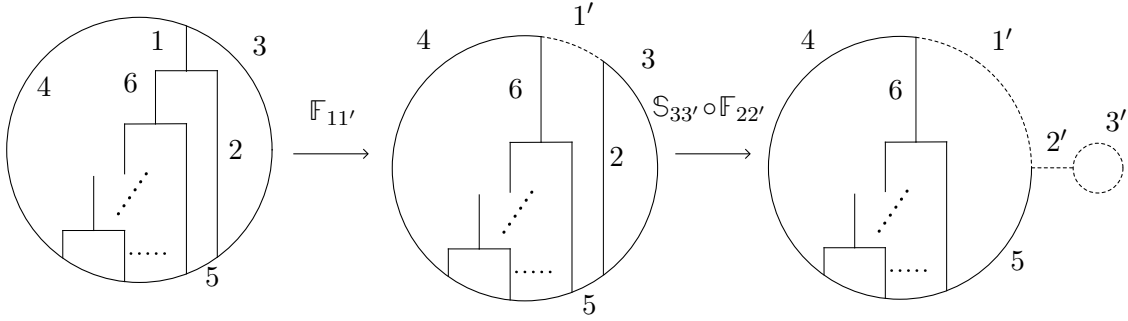
\caption{The sequence of transformations we use to solve for the statistics of the OPE coefficients associated with the skyline channel.}
\label{recursion skyline}
\end{figure}

\begin{figure}
\centering
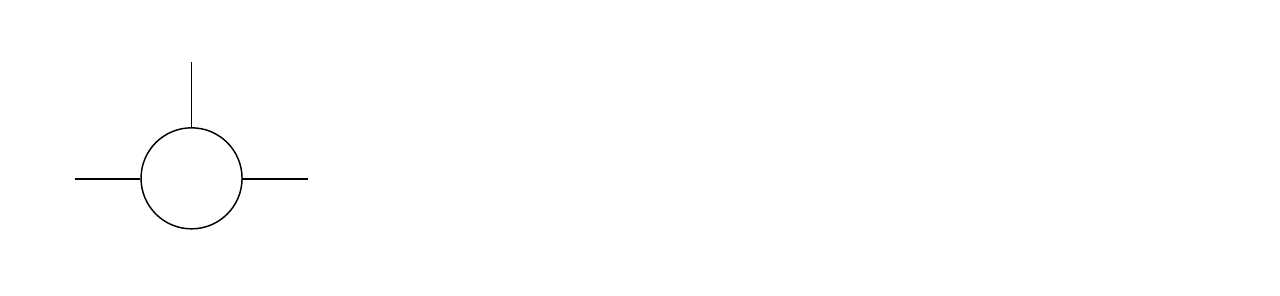
\caption{A summary of the process we use to simplify the skyline channel statistics. This is a relationship between two different OPE densities. Here, $\bar \bbk$ corresponds to the antiholomorphic counterpart of the crossing kernel.  Corrections to this relation are exponentially suppressed at large values of the internal momenta $P_3,\bar P_3$.}
\label{recursion skyline process}
\end{figure}

Our analysis starts with the genus-$g$ partition function in the skyline channel. As in the previous example, we omit the dependence of this function on the moduli and focus on the internal momenta $P_i, \bar P_i$
\begin{equation}
    Z_g = \sum_{\clo_1,\dots,\clo_{3(g-1)}} C_{\{\textnormal{skyline}\}} \clf_S(P_{\clo_i}) \bar \clf_S(\bar P_{\clo_i}) = \int \prod_{i=1}^{3g-3} \frac{dP_i}{2}\frac{d\bar P_i}{2} \rho_S^{(g)}(P_i,\bar P_i)\clf_S(P_i) \bar \clf_S(\bar P_i).
\end{equation}
Figure \ref{recursion skyline} shows the crossing transformations that we will study. The crossing equation that follows from these transformations reads
\begin{equation}
\label{fullrecursion}
\begin{split}
    \rho_S^{(g)} &= \int \frac{dP_1'}{2}\frac{dP_2'}{2}\frac{dP_3'}{2}\frac{d\bar P_1'}{2}\frac{d\bar P_2'}{2}\frac{d\bar P_3'}{2} \bigg(
    \fker{P_1}{P_1'}{P_3}{P_4}{P_6}{P_2}
    \fker{P_2}{P_2'}{P_3}{P_1'}{P_5}{P_3}
    \sker{P_3}{P_3'}{P_2'} \times \bar \bbk
    \\ & \eqspace{3cm} \times \sum_{\clo_1',\clo_2',...} C_{3'3'2'}C_{1'2'5}C_{1'46}C\cdots \left[\delta(P_3'-P_{\clo_3'})\delta(P_2'-P_{\clo_2'})\cdots\right]\bigg)\\
    &= \sum_{\clo_1',\clo_2',\clo_3'} \fker{P_1}{P_{\clo_1'}}{P_3}{P_4}{P_6}{P_2}
    \fker{P_2}{P_{\clo_2'}}{P_3}{P_{\clo_1'}}{P_5}{P_3}
    \sker{P_3}{P_{\clo_3'}}{P_{\clo_2'}}\times \bar \bbk\\
    &\eqspace{3cm} \times C_{3'3'2'} \sum_{\clo_4,\clo_5,\dots}\;\;
     \; C_{1'2'5}C_{1'46}C\dots \left[\delta(P_4-P_{\clo_4})\delta(P_5-P_{\clo_5})\cdots\right].
\end{split}
\end{equation}
Here, $\bar \bbk$ stands for the antiholomorphic counterpart of the crossing kernel and we have made explicit the difference between the discrete  $P_{\clo_i}$ variables in the sum, and the continuum $P_i$ variables in the integrals and delta functions. This time, we have also made explicit that the structure of the spectral densities of OPE coefficients is a sum over delta functions whose weights are OPE coefficients.

As with the genus-three example, the modular kernel $\sker{P_3}{P_{\clo_3'}}{P_{\clo_2'}}$ is exponentially suppressed at large $P_3$. Despite being part of the crossing kernel, the variables $P_2'$ and $P_1'$ are not exponentially suppressed or enhanced. This makes $\clo_{3'} = \bbi$ the dominant contribution in the sum over $\clo_{3'}$. Note that $3'=\bbi$ implies that $2'=\bbi$ and $1'= 5$ via the OPE coefficients $C_{3'3'2'}$ and   $C_{1'2'5}$. After substituting these variables in the first term of equation \rref{fullrecursion}, we are left with the following recursion formula for the genus-$g$ spectral density of OPE coefficients in the skyline channel 
\begin{equation}
\label{Eq:recursionSkyline}
\begin{split}
    \rho_S^{(g)} \approx \fker{1}{5}{3}{4}{6}{2}\fker{2}{\bbi}{3}{5}{5}{3}\sker{3}{\bbi}{\bbi}\times \bar\bbk\;\rho_ S^{(g-1)}.
\end{split}
\end{equation}

Let us make a few comments about equation \rref{Eq:recursionSkyline}. First, although we derived this equation in the context of the skyline channel OPE density, the approach we used to remove one of the heavy lines in the graph of Figure \ref{recursion skyline} is more general. Figure \ref{recursion skyline process} summarizes the process we are using. In the heavy limit, we can always use this sequence of transformations to remove one heavy line from the left diagram in Figure \ref{recursion skyline process}. The resulting density times the three kernels is a good approximation of the previous density up to additive corrections that are exponentially suppressed in the heavy momentum $P_3$. Second, the formula \rref{Eq:recursionSkyline} and the process described in Figure \ref{recursion skyline process} only require us to have heavy momenta so we can use these formulas to study different asymptotic limits. 

In the limit of large dimension but fixed differences, the three kernels simplify to\footnote{In this discussion, we omit the contributions coming from the $\delta_i$ variables since they only affect the order one terms in the final result.} 
\begin{equation}
\label{additivetermsinOPEdensity}
\log \fker{P_1}{P_5}{P_3}{P_4}{P_6}{P_2}\fker{P_2}{\bbi}{P_3}{P_5}{P_5}{P_3}\sker{P_3}{\bbi}{\bbi} = 3P^2 \log \frac{27}{16}+\pi Q P  + \frac{5 Q^2-1}{6}\log P + \clo(1).
\end{equation}
Now, we can use the genus-three OPE density in equation \rref{OPEgen3density}, as our base step to solve the recursion relation in \rref{Eq:recursionSkyline}; we find that 
\begin{equation}
    \log \rho_S^{(g)} \sim 3(g-1)P^2\log \frac{27}{16} + (g+1)\pi Q P + (g-1)\frac{5 Q^2-1}{6}\log P+\textnormal{ (a.c.) }.
\end{equation}
To get the OPE statistics associated with this channel, we have to subtract $3(g-1)$ factors of the entropy. The final result is 
\begin{equation}
\label{skylinestatistics}
   \overline{C_{\{\textnormal{skyline}\}}} \approx \left(\frac{27}{16}\right)^{3\frac{\nn}{2}\Delta}e^{\left(2-5 \frac{\nn}{2}\right)\pi Q(P+\bar P)}(P\bar P)^{\frac{5Q^2-1}{12}\nn},
\end{equation}
% \begin{equation}
% \label{skylinestatistics}
%   \overline{C_{\{\textnormal{skyline}\}}} \approx \left(\frac{27}{16}\right)^{3(g-1)\Delta}e^{(7-5g)\pi Q(P+\bar P)}(P\bar P)^{(g-1)\frac{5Q^2-1}{6}}
% \end{equation}
where we have written our result in terms of the number of OPE coefficients in the average $\nn = 2(g-1)$. This formula is valid for $\nn \geq 2$, or equivalently, $g\geq 2$. At $\nn=2$, the result simplifies to the average value of the squared of OPE coefficients derived in \cite{Collier:2019weq}.

\subsection{The comb channel}
\label{section 3.2}

We now derive a recursion formula for the comb channel OPE statistics. The base step of this recursion formula is the genus-two partition function, this time in the dumbbell channel.

\subsubsection{At genus two}
\begin{figure}
    \centering
    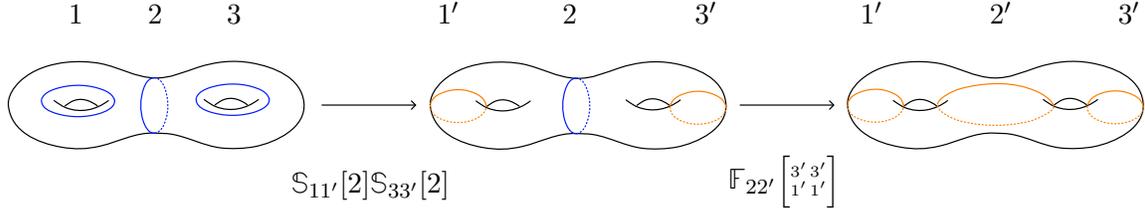
    \caption{The different pair-of-pants decompositions of the genus-two Riemann surface and the crossing transformations that relate them. The leftmost cutting corresponds to the dumbbell channel and the rightmost to the sunset channel.}
    \label{genus2}
\end{figure}

The sequence of transformations we will consider is depicted in  Figure \ref{genus2}. In the dumbbell and sunset channels, the partition function reads 
\begin{equation}
    Z_2 = \sum_{\clo_1\clo_2\clo_3} C_{112}C_{233}\clf_D\clf_D = \sum_{\clo_1'\clo_2'\clo_3'}C_{1'2'3'}C_{1'2'3'} \clf_S\clf_S.
\end{equation}
% The sequence of moves relating the two conformal blocks has two modular kernels and one fusion kernel
% \begin{equation}
%     \clf_S(P_1',P_2',P_3') = \int\frac{dP_1}{2}\frac{dP_2}{2}\frac{dP_3}{2} \sker{1}{1'}{2}\sker{3}{3'}{2}\fker{2}{2'}{3'}{3'}{1'}{1'} \clf_D(P_1,P_2,P_3).
% \end{equation}
The crossing equation relating the two different OPE densities is  
\begin{equation}
    \rho_D(P_i,\bar P_i) = \sum_{\clo_1'\clo_2'\clo_3'} C_{1'2'3'}^2 \;\sker{P_1}{P_{\clo_1'}}{P_2}\sker{P_3}{P_{\clo_3'}}{P_2}\fker{P_2}{P_{\clo_2'}}{P_{\clo_3'}}{P_{\clo_3'}}{P_{\clo_1'}}{P_{\clo_1'}}\times\bar \bbk.
\end{equation}
This time, we are considering a different limit of the modular kernel. Like the fusion kernel, the modular kernel has a rich analytic structure and its behaviour changes depending on the asymptotics. In the large-weight limit with fixed differences, the modular kernel has the following behaviour (see  appendix \ref{appendixmodular}) 
\begin{multline}
\log \sker{P_1}{P_1'}{P_2} = \left(-4\log 2 + \frac{9\log 3}{2}\right)P^2 + \left[\pi\left(Q-2\alpha_1'\right)+ 2\delta_1 \log\frac{27}{16}+\delta_2\log27\right]P\\+\left(\frac{1+7Q^2}{6}-4h_1'\right)\log P + \clo(1), \quad P_1,P_2\gg c, |P_1-P_2|.     
\end{multline}
From this expression, we note that the momenta $P_1'$ and $P_3'$ in the crossing equation are exponentially suppressed at large $P$. The momentum $P_2'$ is, on the other hand, exponentially suppressed by the fusion kernel. Unlike the genus-three partition function, this time the identity contribution vanishes, meaning that our results will depend on the light data of the theory. There are several ways to show this, a particularly simple one is to use the formula \rref{skylinedelta} in the appendix, the result is
\begin{equation}
    \fker{P_2}{\bbi}{\bbi}{\bbi}{\bbi}{\bbi} = \delta(P_2-i Q/2).
\end{equation}
Since we are working in the heavy limit $P_2\gg c$ and the other two terms $\sker{P_1}{\bbi}{P_2}$ and $\sker{P_3}{\bbi}{P_2}$ in the crossing kernel are regular, this first contribution to the sum vanishes. From the next three corrections to the identity kernel in the crossing equation, two of them are zero
\begin{equation}  
\fker{2}{\chi}{\chi}{\chi}{\bbi}{\bbi} = \fker{2}{\chi}{\bbi}{\bbi}{\chi}{\chi} = \delta(P_2-i Q/2)= 0.
\end{equation}
These results follow from equation \rref{combdelta}. Thus, we are left with the relatively simple expression 
\begin{equation}
\label{Eq:comb}
    \rho_D(P_i,\bar P_i) \approx \sker{P_1}{P_\chi}{P_2}\sker{P_3}{P_\chi}{P_2}\fker{P_2}{\bbi}{P_\chi}{P_\chi}{P_\chi}{P_\chi}\times \sker{\bar P_1}{\bar P_\chi}{\bar P_2}\sker{P_3}{\bar P_\chi}{\bar P_2}\fker{\bar P_2}{\bbi}{\bar P_\chi}{\bar P_\chi}{\bar P_\chi}{\bar P_\chi}.
\end{equation}
This result depends on the lightest operator in the theory, regardless of whether it couples or not to the heavy momenta. This formula can be seen as a consequence, via modular invariance, of the OPE data $C_{\chi\chi \bbi}^2 = 1$.

After substituting the asymptotics of the crossing kernels in equation \rref{Eq:comb}, we find
\begin{equation}
\log \rho_D(P_i,\bar P_i) \approx  3P^2 \log \frac{27}{16} +  \pi (3 Q - 4 \alpha_\chi) P + \frac{5Q^2-1}{6}\log P +\textnormal{ (a.c.)}.
\end{equation}
Here, we have set $(\delta_1+\delta_2+\delta_3) = 0$. The final result is
\begin{equation}
\label{combgen2stat}
   \overline{C_{112}C_{233}} \approx \left(\frac{27}{16}\right)^{3\Delta}e^{- \pi(3Q +4\alpha_\chi) P-\pi(3Q +4\bar \alpha_\chi)\bar P}\left(P\bar P\right)^{\frac{5Q^2-1}{6}}.
\end{equation}
Note that the modular and fusion kernels only guarantee an exponential suppression if the lightest operator in the sum, this time corresponding to the lightest primary operator in theory that is not the identity, is in the discrete range ($h_\chi< \frac{c-1}{24}$). If this is not the case, then corrections to equation \rref{combgen2stat} become relevant. It is interesting that this is the expected behaviour in a putative theory of pure gravity where there are no light operators in the spectrum. We will return to this question in the discussion section.

\subsubsection{At higher genus}

\begin{figure}
\centering
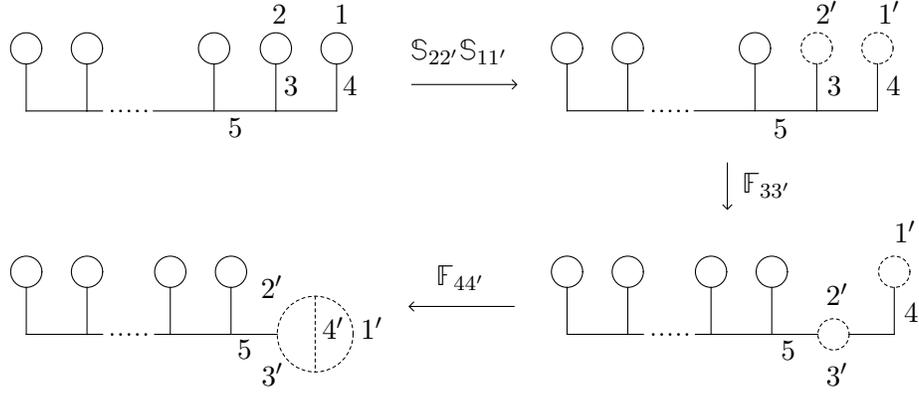
\caption{The sequence of transformations we use to find the statistics of the OPE coefficients in the comb channel.}
\label{recursion comb}
\end{figure}
\begin{figure}
\centering
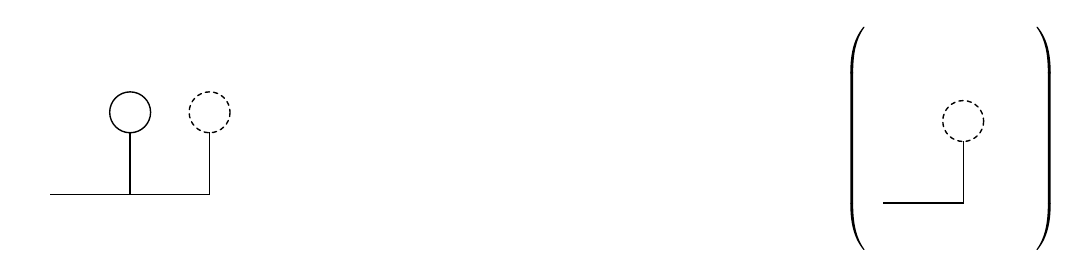
\caption{The general process used to simplify the comb channel. Here, $\bar\bbk$ corresponds to the antiholomorphic counterpart of the crossing kernel. This relationship is valid in the limit where $P_i \gg c, |P_i-P_j|$ and corrections are exponentially suppressed in $P_i$.}
\label{recursion comb process}
\end{figure}

The crossing transformations we will study in this section are described in Figure \ref{recursion comb}. We start by applying a modular transformation to the first loop in the graph 
\begin{equation}
    \begin{split}
        \rho_C^{(g)} &= \int\frac{dP_1'}{2}\frac{d\bar P_1'}{2}\bigg(
        \sker{P_1}{P_1'}{P_4}\sker{\bar P_1}{\bar P_1'}{\bar 4}\sum_{\clo_1',\clo_2,\dots} C_{1'1'4}C_{223}C_{345}C\cdots\left[\delta(P_1'-P_{\clo_1'})\delta(P_2-P_{\clo_2})\cdots\right]\bigg)\\
        &\approx \sker{P_1}{P_\chi}{P_4}\sker{\bar P_1}{\bar P_\chi}{\bar P_4}\sum_{\clo_2,\clo_3,...} C_{\chi\chi 4}C_{223}C_{345}C\cdots\left[\delta(P_2-P_{\clo_2})\cdots\right]\\
        &=: \sker{P_1}{P_\chi}{P_4}\sker{\bar P_1}{\bar P_\chi}{\bar P_4} \rho_{\chi}^{(g)}.
    \end{split}
\end{equation}
In the last line of this equation, we have defined the OPE density $\rho_\chi^{(g)}$. Now, we use the remaining transformations to derive a recursion formula for this density. The result of the transformation is shown in Figure \ref{recursion skyline process}.
\begin{equation}
\label{crossing moves comb}
\begin{split}
    \rho_\chi^{(g)} &= 
    \int \frac{dP_2'}{2}\frac{dP_3'}{2}\frac{dP_4'}{2}
    \frac{d\bar P_2'}{2}\frac{d\bar P_3'}{2}\frac{d\bar P_4'}{2}
    \bigg(
    \sker{P_2}{P_2'}{P_4}
    \fker{P_3}{P_3'}{P_2'}{P_2'}{P_4}{P_5}
    \fker{P_4}{P_4'}{P_\chi}{P_\chi}{P_2'}{P_3'}\times \bar \bbk
    \\
    &\eqspace{3cm}\times \sum_{\clo_2'\clo_3'\dots} C_{\chi 2'4'}C_{\chi 4'3'}C_{2'3'5}C\cdots\left[\delta(P_2'-P_{\clo_2'})\delta(P_3'-P_{\clo_3'})\cdots\right]
    \bigg)\\
    &= \sum_{\clo_2'\clo_3'\clo_4'}
    \sker{P_2}{P_{\clo_2'}}{P_3}
    \fker{P_3}{P_{\clo_3'}}{P_{\clo_2'}}{P_{\clo_2'}}{P_4}{P_5}
    \fker{P_4}{P_{\clo_4'}}{P_\chi}{P_\chi}{P_{\clo_2'}}{P_{\clo_3'}}\times \bar \bbk
    \\
    &\eqspace{3cm}\times  C_{\chi 2'4'}C_{\chi 4'3'}\sum_{\clo_5,\clo_6,\dots}C_{2'3'5}C\cdots\left[\delta(P_5-P_{\clo_5})\delta(P_6-P_{\clo_6})\cdots\right].
\end{split}
\end{equation}
In the sum, $P_2'$ and $P_4'$ are suppressed by the modular and fusion kernels, respectively. As in the genus-two partition function, the identity contribution vanishes. Intuitively, this happens because in the second step of Figure \ref{recursion comb} the OPE coefficient $C_{2'2'3}$ is zero if we take $2' = \bbi$. Moreover, the third diagram includes the OPE coefficient $C_{2'45}$ which is zero unless $P_4=P_5$. These OPE coefficients do not appear in the final configuration shown in equation \rref{crossing moves comb}, but their data are kept in the crossing kernels. We can extract this information step by step. 

Setting $2' = \bbi$ in equation \rref{crossing moves comb}, implies via the final OPE, that $4' = \chi$ and $3' = 5$. So we have from one of the fusion kernels  
\begin{equation}
    \fker{4}{\chi}{\bbi}{5}{\chi}{\chi} = \delta(P_4-P_5);
\end{equation}
recovering one of the constraints. The other fusion kernel then reads, 
\begin{equation}
    \fker{3}{5}{\bbi}{\bbi}{5}{5}=0.
\end{equation}
This corresponds to the vanishing of the OPE coefficient  $C_{2'2'3}$ as we take $2'=\bbi$. 

The first nonzero term in the sum in equation \rref{crossing moves comb} happens when $2'=\chi$ and $P_4' = \bbi$. This sets, via the OPE coefficient $C_{\chi 4' 3'}$, $3' = \chi$. Meaning that, to leading order, we have the following recursion formula 
\begin{equation}
    \rho_\chi^{(g)} \approx \sker{2}{\chi}{3}
    \fker{3}{\chi}{\chi}{\chi}{4}{5}
    \fker{4}{\bbi}{\chi}{\chi}{\chi}{\chi} \times \bar \bbk \; \rho_{\chi}^{(g-1)}.
\end{equation}
Corrections to this equation are exponentially suppressed in $P$. In terms of comb-channel OPE statistics, this equation reads
\begin{equation}
\label{combrecursion}
    \rho_C^{(g)} \approx 
    \sker{P_2}{P_\chi}{P_3}
    \fker{P_3}{P_\chi}{P_\chi}{P_\chi}{P_4}{P_5}
    \fker{P_4}{\bbi}{P_\chi}{P_\chi}{P_\chi}{P_\chi} \times \bar \bbk \;\rho_C^{(g-1)} \,.
\end{equation}
The base step of this recursion are the statistics of the genus-two comb channel (i.e. the dumbbell channel). Recall that we are working in the limit $P_i - \delta_i = P\rightarrow \infty$. Here, the crossing kernels simplify to\footnote{As with the skyline channel, we omit the contributions coming from the $\delta_i$ variables, since they only contribute to the order-one terms in the final result.} 
\begin{equation}
    \log \sker{P_2}{P_\chi}{P_3}
    \fker{P_3}{P_\chi}{P_\chi}{P_\chi}{P_4}{P_5}
    \fker{P_4}{\bbi}{P_\chi}{P_\chi}{P_\chi}{P_\chi}=3 P^2\log\frac{27}{16}+\frac{\pi}{2}\left(3Q - 4 \alpha_\chi\right)P+\frac{5Q^2-1}{6}\log P + \clo(1) \,,
\end{equation}
where we have used the asymptotics of the fusion kernel with two light external operators discussed in appendix \ref{appendixfusion}. Solving the recursion formula \rref{combrecursion} yields
\begin{equation}
    \log \rho_C^{(g)} \sim 3(g-1)P^2 \log \frac{27}{16}+\frac{1}{2}g \pi (3Q-4\alpha_\chi)P  + (g-1)\frac{5Q^2-1}{6}\log P + \textnormal{ (a.c.)}.
\end{equation}
Dividing by the corresponding densities, we arrive at the following formula for the average value of OPE coefficients in the comb channel configuration
\begin{equation}
\label{Eq:combstatistics}
    \overline{C_{\{\textnormal{comb}\}}} \approx \left(\frac{27}{16}\right)^{3 \frac{\nn}{2}\Delta}e^{\left[\frac{3}{2}\left(1-3\frac{\nn}{2}\right)-2\left(\frac{\nn}{2}+1\right)\frac{\alpha_\chi}{Q}\right]\pi QP+\left[\frac{3}{2}\left(1-3\frac{\nn}{2}\right)-2\left(\frac{\nn}{2}+1\right)\frac{\bar \alpha_\chi}{Q}\right]\pi Q\bar P}\left(P\bar P\right)^{ \frac{5Q^2-1}{12}\nn},
\end{equation}
% \begin{multline}
% \label{Eq:combstatistics}
%     \overline{C_{\{\textnormal{comb}\}}} \approx \left(\frac{27}{16}\right)^{3(g-1)\Delta}
%     \exp{-\frac{\pi}{2}\left[3(3g-4)Q+4g\alpha_\chi\right]P
%     -\frac{\pi}{2}\left[3(3g-4)Q+4g\bar \alpha_\chi\right]\bar P}\\
%     \times \exp{(g-1)\frac{5Q^2-1}{6}\left(\log P+\log \bar P\right)}
% \end{multline}
where we have expressed the result in terms of the number of OPE coefficients in the average $\nn$. This time, the contribution of two extra OPE coefficients to the average corresponds to an extra factor of $e^{-\frac{9}{2}\pi Q(P+\bar{P})}$, which yields a suppression of $e^{-\frac{9S}{4}}$ plus non-universal corrections. Comparing this result with the skyline channel, where two additional coefficients correspond to a suppression of $e^{-\frac{5 S}{2}}$, we see that the OPE statistics not only depend on the topology of the Riemann surface in the partition function, but also on the topology of their trivalent diagram. 

\subsection{Estimating other channels}
\label{section 3.3}

\begin{figure}
\centering
%% Creator: Inkscape 1.0 (4035a4fb49, 2020-05-01), www.inkscape.org
%% PDF/EPS/PS + LaTeX output extension by Johan Engelen, 2010
%% Accompanies image file 'otherfamilies.pdf' (pdf, eps, ps)
%%
%% To include the image in your LaTeX document, write
%%   \input{<filename>.pdf_tex}
%%  instead of
%%   \includegraphics{<filename>.pdf}
%% To scale the image, write
%%   \def\svgwidth{<desired width>}
%%   \input{<filename>.pdf_tex}
%%  instead of
%%   \includegraphics[width=<desired width>]{<filename>.pdf}
%%
%% Images with a different path to the parent latex file can
%% be accessed with the `import' package (which may need to be
%% installed) using
%%   \usepackage{import}
%% in the preamble, and then including the image with
%%   \import{<path to file>}{<filename>.pdf_tex}
%% Alternatively, one can specify
%%   \graphicspath{{<path to file>/}}
%% 
%% For more information, please see info/svg-inkscape on CTAN:
%%   http://tug.ctan.org/tex-archive/info/svg-inkscape
%%
\begingroup%
  \makeatletter%
  \providecommand\color[2][]{%
    \errmessage{(Inkscape) Color is used for the text in Inkscape, but the package 'color.sty' is not loaded}%
    \renewcommand\color[2][]{}%
  }%
  \providecommand\transparent[1]{%
    \errmessage{(Inkscape) Transparency is used (non-zero) for the text in Inkscape, but the package 'transparent.sty' is not loaded}%
    \renewcommand\transparent[1]{}%
  }%
  \providecommand\rotatebox[2]{#2}%
  \newcommand*\fsize{\dimexpr\f@size pt\relax}%
  \newcommand*\lineheight[1]{\fontsize{\fsize}{#1\fsize}\selectfont}%
  \ifx\svgwidth\undefined%
    \setlength{\unitlength}{311.81102362bp}%
    \ifx\svgscale\undefined%
      \relax%
    \else%
      \setlength{\unitlength}{\unitlength * \real{\svgscale}}%
    \fi%
  \else%
    \setlength{\unitlength}{\svgwidth}%
  \fi%
  \global\let\svgwidth\undefined%
  \global\let\svgscale\undefined%
  \makeatother%
  \begin{picture}(1,0.22727273)%
    \lineheight{1}%
    \setlength\tabcolsep{0pt}%
    \put(0,0){\includegraphics[width=\unitlength,page=1]{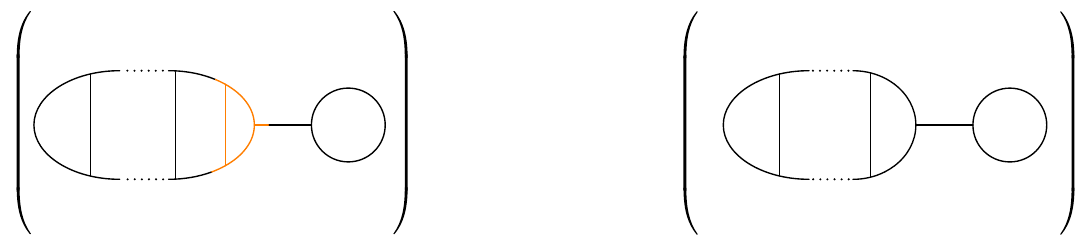}}%
    \put(0.42486449,0.10258417){\makebox(0,0)[lt]{\lineheight{1.25}\smash{\begin{tabular}[t]{l}$ \sim \; \bbs\circ\bbf\circ\bbf$\end{tabular}}}}%
  \end{picture}%
\endgroup%

\caption{An example of another channel that can be simplified using the skyline and comb reductions.}
\label{other recursion}
\end{figure}

We can use the transformations we have studied so far to solve different OPE configurations. The relationships described in Figures \ref{recursion skyline process} and \ref{recursion comb process} are valid regardless of the specific channel. The skyline recursion in Figure \ref{recursion skyline process} only requires the heavy momenta $P_i$ and is valid in different asymptotic limits. The comb recursion in Figure \ref{recursion comb process} is valid when the differences between the momenta $P_i$ are held fixed. 

As an example of another channel that can be simplified using these transformations, consider the channel depicted in Figure \ref{other recursion}. This channel can be solved by repeatedly using the skyline recursion formula. Solving for the statistics of this channel in the heavy limit where the differences between the momenta are held fixed is straightforward. We start by counting how many loops we need to contract from the diagram in order to get the genus-two dumbbell channel; this channel corresponds to the  base step of this family of statistics. For each loop, we contract we have to multiply the base-step OPE density by the factor given in equation \rref{additivetermsinOPEdensity}. The last step is to divide by as many factors of the density $\rho_0(P,\bar P)$ as heavy lines in the diagram, for an $n$-punctured Riemann surface of genus $g$, this number is $2g+n-2$. The end result for this channel is
\begin{equation} \label{mixedchannel}
   \overline{C_{%% Creator: Inkscape 1.0 (4035a4fb49, 2020-05-01), www.inkscape.org
%% PDF/EPS/PS + LaTeX output extension by Johan Engelen, 2010
%% Accompanies image file 'OPEcoef.pdf' (pdf, eps, ps)
%%
%% To include the image in your LaTeX document, write
%%   \input{<filename>.pdf_tex}
%%  instead of
%%   \includegraphics{<filename>.pdf}
%% To scale the image, write
%%   \def\svgwidth{<desired width>}
%%   \input{<filename>.pdf_tex}
%%  instead of
%%   \includegraphics[width=<desired width>]{<filename>.pdf}
%%
%% Images with a different path to the parent latex file can
%% be accessed with the `import' package (which may need to be
%% installed) using
%%   \usepackage{import}
%% in the preamble, and then including the image with
%%   \import{<path to file>}{<filename>.pdf_tex}
%% Alternatively, one can specify
%%   \graphicspath{{<path to file>/}}
%% 
%% For more information, please see info/svg-inkscape on CTAN:
%%   http://tug.ctan.org/tex-archive/info/svg-inkscape
%%
\begingroup%
  \makeatletter%
  \providecommand\color[2][]{%
    \errmessage{(Inkscape) Color is used for the text in Inkscape, but the package 'color.sty' is not loaded}%
    \renewcommand\color[2][]{}%
  }%
  \providecommand\transparent[1]{%
    \errmessage{(Inkscape) Transparency is used (non-zero) for the text in Inkscape, but the package 'transparent.sty' is not loaded}%
    \renewcommand\transparent[1]{}%
  }%
  \providecommand\rotatebox[2]{#2}%
  \newcommand*\fsize{\dimexpr\f@size pt\relax}%
  \newcommand*\lineheight[1]{\fontsize{\fsize}{#1\fsize}\selectfont}%
  \ifx\svgwidth\undefined%
    \setlength{\unitlength}{32.33522911bp}%
    \ifx\svgscale\undefined%
      \relax%
    \else%
      \setlength{\unitlength}{\unitlength * \real{\svgscale}}%
    \fi%
  \else%
    \setlength{\unitlength}{\svgwidth}%
  \fi%
  \global\let\svgwidth\undefined%
  \global\let\svgscale\undefined%
  \makeatother%
  \begin{picture}(1,0.27263377)%
    \lineheight{1}%
    \setlength\tabcolsep{0pt}%
    \put(0,0){\includegraphics[width=\unitlength,page=1]{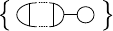}}%
  \end{picture}%
\endgroup%
}} \approx 
   \left(\frac{27}{16}\right)^{3(g-1)\Delta}e^{\pi\left[(7-5g) Q -2\alpha_\chi\right]P+\pi\left[(7-5g) Q -2\bar \alpha_\chi\right]\bar P}\left(P \bar P\right)^{ (g-1)\frac{5Q^2-1}{6}}.
\end{equation}

In general, we can use these methods to simplify densities with three heavy lines attached to one loop or two loops attached to a single heavy line. Note that this method is valid even if the observable is not just a partition function, but also a correlation function since the crossing kernels act locally on a few heavy lines.

\section{Euclidean wormholes and typicality}

We now discuss the implications of our results for holography and Euclidean wormholes. In \cite{Belin:2020hea}, the proposed interpretation of Euclidean wormholes was that it corresponds to treating the OPE coefficients of the CFT as random variables rather than using the true microscopics. Indeed, a low energy observer\footnote{By this, we mean any physical quantity that can be calculated with the low-energy effective action of gravity, i.e. by the Einstein-Hilbert action.} cannot distinguish between the black hole microstates (or precisely compute their energy or OPE coefficients), but can nevertheless compute the distribution of such microstates over sufficiently large energy windows. We now explain the interpretation of the results presented in this paper from that point of view.

To be concrete, we will focus our attention on computing the product of two genus-two partition functions, and revisit the genus-two wormhole described in \cite{Maldacena:2004rf}. Consider a CFT$_2$ living on two disconnected copies of a genus-two surface. The partition function of this theory is
\begin{equation} \label{twoproducts}
    Z_{g=2\times g=2}(\beta_i) = \left(Z_{g=2}(\beta_i)\right)^2 = \sum_{\clo_1\cdots\clo_6} C_{123}^2C_{456}^2 |\clf_S\left(\{h_{\clo_{123}}\},\{\beta_{i}\}\right)|^2|\clf_S\left(\{ h_{\clo_{456}}\},\{ \beta_i\}\right)|^2 \prod_{i=1}^6 q_i^{\Delta_i} \,,
\end{equation}
where for simplicity, we study the case where the two genus-two surfaces share the same moduli $\beta_i$, where $q_i=e^{-2\pi \beta_i}$. The general prescription for computing this object in the bulk is to find the solutions of the gravitational equations of motion with appropriate boundary conditions:
\begin{equation}
    Z_{g=2\times g=2}^{\textnormal{grav}}(\beta_i) = \int_{\partial\mathcal{\M} = \Sigma(\beta_i)\times\Sigma(\beta_i)} Dg\; e^{-\EH(g)} = \sum_{g^*}e^{- \EH(g^*) + c^{-1} \EH^{(1)}(g^*)+\cdots},
\end{equation}
$\Sigma(\beta_i)\times \Sigma(\beta_i)$ corresponds to the two disconnected genus-two surfaces at the boundary of AdS$_3$ and $\EH$ is the Einstein-Hilbert action
\begin{equation}
\EH = \frac{1}{16\pi G_N}\int d^3x \sqrt{g}\left(R + \frac{2}{\ell^2}\right),
\end{equation}
where $\ell$ is the AdS$_3$ radius. The central charge of the dual CFT and the AdS$_3$ radius are related via \cite{Brown:1986nw}
\begin{equation}
c = \frac{3\ell}{2G_N} \,.
\end{equation}
The sum over $g^*$ is over all classical solutions to the equations of motion, and  $\EH^{(1)}$ is the one-loop correction to the classical action.\footnote{There seem to be cases where including non-saddle geometries in the path integral over metrics still yields sensible results - see for example \cite{Saad:2018bqo,Cotler:2020lxj}. Here, we will only discuss the path integral in terms of saddle-points, but it would be interesting to understand the role of non-saddle geometries better.} The terms inside the ellipsis denote the infinite series of subleading perturbative corrections. 

The sum over geometries includes disconnected solutions known as handlebodies, as well as Euclidean wormholes connecting the two asymptotic boundaries. The simplest connected geometry is the genus-two wormhole, a locally AdS$_3$ metric whose metric is
\be
\frac{ds^2}{\ell^2}=d\rho^2 +\cosh^2\rho d\Sigma_g^2 \,,
\ee
where $\Sigma_g$ is the constant curvature metric on the genus-two surface. The on-shell action of this geometry vanishes (see \cite{Maxfield:2016mwh}).

Connected contributions like the genus-two wormhole seem to destroy the factorization of the CFT partition function on two disconnected manifolds, a puzzle raised in \cite{Maldacena:2004rf}. The proposal put forward in \cite{Belin:2020hea} states that gravity is not computing the exact partition function of its dual CFT \rref{twoproducts}; it only computes an approximation
\begin{equation}
\zg(\beta_i) = \sum_{\clo_1\cdots\clo_6}\overline{C_{123}^2C_{456}^2} | \clf_S \bar  \clf_S |^2 \prod_{i=1}^6 q_i^{\Delta_i} \,.
\end{equation}
Note that this is \textit{not} the exact and microscopic CFT answer, which obviously factorizes. We have replaced microscopic OPE coefficients with their averaged value in a distributional sense. One can then compute the connected contribution from the CFT point of view
\begin{equation}
\zw = \zg -\left(Z^{\textnormal{grav}}_{g=2}\right)^2 = \sum_{\clo_1\cdots\clo_6}    \left(\overline{C_{123}^2C_{456}^2} - \overline{C_{123}^2}\;\;\overline{C_{456}^2}\right)| \clf_S\bar \clf_S|^2 \prod_{i=1}^6 q_i^{\Delta_i}\,.
\end{equation}
 At high temperatures $\beta_i\to0$, the sum is dominated by states of very high energy and we can use our asymptotic formulas for the averaged value of OPE coefficients. Taking into account the Gaussian part of the statistical distribution, \cite{Belin:2020hea} found that the on-shell action of the genus-two wormhole was correctly reproduced. 

The question we would now like to ask is how the non-Gaussianities computed in this paper refine this computation. To do so, we will introduce a generating function for the statistical distribution of OPE coefficients, which can be justified by assuming certain typicality properties of heavy states in CFTs.

\subsection{Typicality and a generating function}

While the ETH is still a conjecture for chaotic quantum systems, it can be ``derived" by assuming that simple operators are essentially random operators inside a given microcanonical window (see for example \cite{DAlessio:2016rwt}). In other words, the change of basis between the eigenbasis of the operator and that of the Hamiltonian is a random unitary.\footnote{In the ergodic limit, i.e. if we restrict to matrix elements of energy eigenstates separated by only a few units of mean level spacing, then one can derive that simple operators are random operators using an EFT of quantum chaos \cite{Altland:2020ccq,Altland:2021rqn}.} For example, we have
\be
\braket{m|O|n}=\sum_{i}O_i \braket{m|i}\braket{i|n} \,.
\ee
If we assume that the change of basis is random, we can replace the overlaps $\braket{m|i}$ by unitaries, which should be averaged over. We then find
\be \label{1ptaverage}
\bra{m}O\ket{n}=\sum_{i}O_i \overline{U_{mi}U^*_{in}}=\delta_{m,n} \overline{O}  \,,
\ee
where $\overline{O}=e^{-S} \sum_{i}O_i$. By assuming that the change of basis is a random unitary, we have derived the diagonal part of ETH. One can proceed in a similar fashion for the variance of $O$, and find the correction that scales as $e^{-S/2}$ \cite{DAlessio:2016rwt}. In fact, one can proceed to study $k$-point correlation function using typicality, which is precisely the venture undertaken by Foini and Kurchan to study the higher-point moments of matrix elements \cite{Foini:2018sdb}. A slight complication that appears within this framework (which will also be relevant for the HHH OPE coefficients) is that one must be careful whether there are repeated indices or not. Typicality also determines the average over repeated indices, but with a slightly more complicated structure. For example, we have \cite{Foini:2018sdb}
\be
\overline{O_{ij}O_{ji}O_{ik}O_{ki}} = \overline{O_{ij}O_{ji}}\ \overline{O_{ik}O_{ki}} + \overline{O_{ij}O_{jm}O_{mk}O_{ki}}\Big|_{m=i} \,.
\ee
For OPE coefficients with three heavy indices, the notion of a random operator acting on the microcanonical Hilbert space is much less clear.\footnote{Note that CFTs have an infinite-dimensional Hilbert space, so it is important to apply the analogy with quantum mechanics described above within a single microcanonical window. The Hamiltonian should not be viewed as an infinite-dimensional random matrix, but rather as a matrix split into blocks of microcanonical windows, where each block resembles a random matrix. Moreover, this analogy also neglects the contribution of descendants which are clearly not random.} However, one can still assume a form of typicality where we take all three indices of the OPE coefficients to be randomized with unitaries. One can even give a physical interpretation of HHH OPE coefficients as operators acting on a tripled Hilbert space  \cite{Belin:2021ibv}. 

Here, instead of studying a typicality-based average of OPE coefficients and carefully track indices that are repeated or not, we will introduce a generating function that keeps track of this in a built-in way. Consider the following generating function\footnote{Note that we have not included ``double-trace" type terms in this generating function. By double-trace, we mean terms that include two disconnected sets of indices that are traced over. We comment on this in the discussion. }
\bea \label{generatingfunction}
\mathcal{Z}(J_{abc})&=&\exp\Big[f_1(\Delta) J_{abc}J^{abc}+f_2(\Delta) J^a_{\;\;ab} J_{\;\;\;c}^{bc}+\sum_{i=1}^{5}g_i(\Delta)JJJJ|_{\textrm{i-type contraction}}\Big] \,,
\eea
where the $i$-type contractions sums over are all possible connected ways to contract the 3-tensor $J$. Note that these are in one-to-one correspondence with the 5 possible channels to decompose a genus-three Riemann surface.\footnote{It is important to emphasize that there is no new information in this generating functional compared to asymptotic formulas of OPE coefficients. It is simply a concise and elegant way to encode the known information, which automatically keeps track of all index structures, including potential repetitions.} Because the OPE coefficients are symmetric under the exchange of all 3 indices, one should take $J$ to be symmetric. 
%To be concise, we omit the sum over permutations from \rref{generatingfunction} and focus solely on the different topologies of contractions, but the sum should be understood as implicit. 
Note that a generating function for Haar-random states and operator matrix elements has appeared before in \cite{Pollack:2020gfa}. While different, the spirit of our generating function certainly shares the same flavour as theirs.

The expression \rref{generatingfunction} is then simply interpreted as the generating function for the statistical distribution of OPE coefficients
\be
\overline{C...C} = \frac{\delta}{\delta J} ... \frac{\delta}{\delta J}\mathcal{Z}(J) \Big|_{J=0} \,.
\ee
Here, the only input of typicality we have used is that the generating function should be constructed from objects that are invariant under the unitary group. Therefore, the terms appearing in the exponent of $\mathcal{Z}$ are the most general terms involving polynomials of $J$ such that all indices are contracted in a unitary-invariant way. The coefficients of the individual terms in the exponent (i.e. the functions $f_i,g_i$) are not determined directly by typicality.\footnote{Note that we have written them as a function of a single energy assuming all states are drawn from the same microcanonical window, but of course, the functions can depend on multiple energy bins as well as energy differences. We use the notation $f(\Delta)$ for simplicity, but it should be understood that these are functions of more variables.} For ETH, these functions would depend on microcanonical traces of the operator and its powers. For example, one such function would be $\overline{O}$ in \rref{1ptaverage}, and other functions can be determined by comparing the prediction of the generating function from the computation of correlation functions.

Unsurprisingly, the way to fix these functions in our case is from the genus-$g$ partition functions. Note that there is always an equivalent number of index contractions of $k$ $J$s and number of inequivalent decompositions of a genus-$g$ surface with $k=2g-2$. Thus one can completely fix the generating function by knowing all the asymptotic formulas for OPE coefficients. In this paper, we have not computed all of these functions at genus-three, but we will nonetheless use the ones we have computed to probe the product of two genus-two partition functions \rref{twoproducts}.\footnote{Note that all contributions enter with positive signs, so the different saddles do not cancel one another. Therefore, the saddles we do compute give a non-vanishing part of the final answer, but the complete set of saddles remains to be computed.} In particular, we have
\bea
g_{\textrm{skyline}}(\Delta)&=&\left(\frac{27}{16}\right)^{6\Delta}e^{-8\pi\sqrt{\frac{c-1}{3}\Delta}}\Delta^{\frac{5c-11}{18}} \\
g_{\textrm{comb}}(\Delta)&=&\left(\frac{27}{16}\right)^{6\Delta}e^{\left[-\frac{15}{2}\pi \sqrt{\frac{c-1}{3}}-\sqrt{18}\pi (\alpha_\chi + \bar \alpha_\chi )\right]\sqrt{\Delta}}\Delta^{\frac{5c-11}{18}} \,.
\eea
where $\chi$ is the lightest operator in the theory.

It is worth comparing results derived from the generating function to the notation adopted in \cite{Belin:2020hea} for the statistical distribution of OPE coefficients. For the Gaussian part of the distribution, the generating function would yield
\be
\overline{C_{abc}C_{def}^*}=f_1(\Delta)\delta_{a,d}\delta_{b,e}\delta_{c,f} +f_2(\Delta) \delta_{a,b}\delta_{c,d}\delta_{e,f}+\textrm{permutations} \,.
\ee
This is exactly the same form as the postulated ansatz of \cite{Belin:2020hea}. The function $f_1$ is given by the asymptotic formula for $|C_{123}|^2$ \cite{Collier:2019weq}
\be
f_1(\Delta)\approx\left(\frac{27}{16}\right)^{3\Delta} e^{-3\pi\sqrt{\frac{c}{3}\Delta}} \,,
\ee
while we derived an asymptotic formula for $f_2$ in this paper from the dumbbell channel at genus-two
\begin{equation}
   f_2(\Delta) \approx \left(\frac{27}{16}\right)^{3\Delta}e^{- 3\pi\sqrt{\frac{c}{3}\Delta}-\frac{4\pi}{\sqrt{2}}(\alpha_\chi+\bar{\alpha}_\chi)\sqrt{\Delta}}.
\end{equation}
We have already taken the large $c$ limit in both of these expressions. At the quartic level, we have partial information and we can write
\bea \label{genus3statdistribution}
\overline{C_{abc}C_{def}^*C_{ghi}C_{jkl}^*} &\supset& g_{\textrm{skyline}}(\Delta) \left[\delta_{a,d}\delta_{b,g}\delta_{c,j}\delta_{e,h}\delta_{f,k}\delta_{i,l}+\textrm{permutations}\right] \notag \\
&+&g_{\textrm{comb}}(\Delta)\left[\delta_{a,b}\delta_{d,e}\delta_{k,l}\delta_{c,g}\delta_{f,h}\delta_{i,j}+\textrm{permutations}\right] \,.
\eea
We will now proceed to study the contributions of these terms for the product of genus-two partition functions.

\subsection{Euclidean wormholes}

We are now ready to consider the various contributions that appear in the product of two genus-two partition functions from the generating function $\mathcal{Z}(J)$. Let us start from the Gaussian data, as was considered in \cite{Belin:2020hea}. This is given by considering the following two contributions\footnote{There are many other contributions we are not writing, either coming from different combinations of Kronecker deltas or coming from terms proportional to $f_1 f_2$ or $f_2^2$. It is easy to see that these terms are simply further exponentially suppressed compared to the terms we kept.}
\be \label{twocontractions}
\overline{C_{abc}C_{def}^*C_{ghi}C_{jkl}^*}\supset f_1(\Delta)^2\left[\delta_{a,d}\delta_{b,e}\delta_{c,f}\delta_{g,j}\delta_{h,k}\delta_{i,l}+\delta_{a,j}\delta_{b,k}\delta_{c,l}\delta_{d,g}\delta_{e,h}\delta_{f,i}\right] \,.
\ee
Both contributions are Gaussian wick contractions, but they will lead to wildly different answers when inserted in the product of genus-two partition functions due to their index nature. From this, we can compute the product of the two genus-two partition functions and find
\bea \label{gaussianresult}
(Z_{g=2})^2\Big|_{\textrm{Gaussian}} &\sim& \sum_{\Delta} \rho(\Delta)^6 \delta_{a,d}\delta_{b,e}\delta_{c,f}\delta_{g,j}\delta_{h,k}\delta_{i,l} \overline{C_{abc}C_{def}^*C_{ghi}C_{jkl}}  | \clf_S \bar  \clf_S |^2 e^{-6\beta \Delta} \notag \\
&\sim& e^{\frac{c}{2}\frac{\pi^2}{\beta}} + \sum_{\Delta} e^{-6\beta \Delta} \,.
\eea
We have used the expression of the genus-two conformal blocks \cite{Cardy:2017qhl} and the first term corresponds to the first Wick contraction in \rref{twocontractions}, while the second term is the second contraction. We see that the second term does not have a saddle-point which yields an answer exponentially large in the central charge. In \cite{Belin:2020hea}, it was observed that this matches the fact that the genus-two wormhole has a vanishing on-shell action.

With this technology in mind, we can now add in the quartic terms coming from the non-Gaussianities in $\mathcal{Z}(J)$. Using \rref{genus3statdistribution}, we find
\bea
(Z_{g=2})^2\Big|_{\textrm{skyline}}&\approx&\sum_{\Delta}e^{8 \pi \sqrt{\frac{c}{3}\Delta}} e^{-8\pi\sqrt{\frac{c}{3}\Delta}} e^{-6\beta \Delta}=\sum_{\Delta}e^{-6\beta \Delta}  \\
(Z_{g=2})^2\Big|_{\textrm{comb}}&\approx& \sum_{\Delta}e^{6\pi \sqrt{\frac{c}{3}\Delta}} e^{\left[-\frac{15}{2}\pi \sqrt{\frac{c}{3}}-\sqrt{18}\pi (\alpha_\chi + \bar \alpha_\chi )\right]\sqrt{\Delta}}e^{-6\beta \Delta}  \notag \\
&\approx& \sum_{\Delta}e^{\left[-\frac{3}{2}\pi \sqrt{\frac{c}{3}}-\sqrt{18}\pi (\alpha_\chi + \bar \alpha_\chi )\right]\sqrt{\Delta}}e^{-6\beta \Delta} \,.
\eea
We see that the comb contribution is more subleading than the Gaussian contribution, and is far from being able to give a large saddle-point contribution. This is true irrespective of the dimension of the lightest operator in the theory. On the other hand, the skyline contribution is just on the threshold, much like the second Wick contraction of the Gaussian distribution. This contribution is thus important: if one wanted to match one-loop determinants around the wormhole geometry, it would be important to take into account the skyline  channel contribution as it is just as big as the second term in \rref{gaussianresult}.

\subsubsection*{The genus-two wormhole in the dumbbell channel}

So far, we have focused on the contribution of the quartic vertices coming from the genus-three partition function in the square of the genus-two partition function, itself expanded in the sunset channel (and in the limit of small moduli). Here, we would like to study the contribution of the quartic vertices for the square of the genus-two partition function, this time in the dumbbell channel. Note that while we can of course expand a genus-two partition function in any channel we want, taking the small moduli limit in one channel probes very different physics than the small moduli limit of another.

The first obstacle we face is that the genus-two conformal blocks are not known (at least to the best of our knowledge) in the dumbbell channel. We will thus have to make a guess for the scaling of conformal blocks, and leave a more detailed analysis for future work. The simple assumption we will make is that the scaling of the conformal blocks in the dumbbell channel is
\be \label{expgrowingfact}
\clf_{\textrm{dumbbell}}\bar \clf_{\textrm{dumbbell}} \sim \left(\frac{16}{27}\right)^{3\Delta} \,,
\ee
namely that the blocks scale in the same way as in the sunset channel. This is only a mild assumption for two reasons. This scaling depends on the normalization for OPE coefficients, and appears only because we decided to normalize the OPE coefficients such that it measures the three-point function of operators inserted at 0, 1 and $\infty$. It is natural that this normalization choice affects all blocks in a uniform fashion. Moreover, the exponentially growing factor of the type $(27/16)^{3\Delta}$ in the asymptotics of the OPE coefficients would prevent convergence of the partition functions if it is not cancelled against the blocks. It would of course still be worth checking this explicitly.

From this, we can estimate the genus-two partition function in the dumbbell channel. 
\be \label{dumbbellpartfunc}
Z_{\textrm{dumbbell}}\approx\sum_{\Delta} \rho(\Delta)^3 e^{- 3\pi\sqrt{\frac{c}{3}\Delta}-\frac{4\pi}{\sqrt{2}}(\alpha_\chi+\bar{\alpha}_\chi)\sqrt{\Delta}}e^{-3\beta\Delta }\approx e^{\frac{c+8\Delta_\chi}{4} \frac{\pi^2}{\beta}} \, ,
\ee
where we have assumed $\Delta_\chi\ll c$, used the Cardy formula and computed the sum over $\Delta$ by saddle-point. We comment on the fact that the lightest non-trivial operator appears here in the discussion section.
Similarly, one can use the Gaussian part of the random ansatz to find
\be
(Z_{\textrm{dumbbell}})^2\Big|_{\textrm{Gaussian}} \approx e^{\frac{c+8\Delta_\chi}{2} \frac{\pi^2}{\beta}} +\sum_{\Delta}e^{-\frac{8\pi}{\sqrt{2}}(\alpha_\chi+\bar{\alpha}_\chi)\sqrt{\Delta}}e^{-6\beta\Delta} \,.
\ee
Similar to the sunset case, we find a contribution which is the square of the genus-two partition function plus a second term which does not lead no a large saddle. In fact, in the dumbbell case, we see that the situation even gets worsened by the correction coming from the lightest operator in the theory and we are no longer at the threshold of having a large saddle.

One can now take into account the contribution of the comb quartic term. We find
\bea
(Z_{\textrm{dumbbell}})^2\Big|_{\textrm{comb}}&\approx& \sum_\Delta \rho(\Delta)^5 e^{\left[-\frac{15}{2}\pi \sqrt{\frac{c}{3}}-\sqrt{18}\pi (\alpha_\chi + \bar \alpha_\chi )\right]\sqrt{\Delta}}e^{-6\beta\Delta} \notag \\ 
&\approx&\sum_{\Delta}e^{\left[\frac{5}{2}\pi \sqrt{\frac{c}{3}}-\sqrt{18}\pi (\alpha_\chi + \bar \alpha_\chi )\right]\sqrt{\Delta}}e^{-6\beta\Delta} \notag \\
&\approx&e^{\frac{25c-360\Delta_\chi}{288} \frac{\pi^2}{\beta}}
\,.
\eea
We now see a remarkable feature! The sum over conformal dimensions can have a non-trivial saddle point which yields an answer exponentially large in $c$. In the gravitational language, this should be a new saddle-point geometry with a non-vanishing on-shell action. This saddle-point only exists provided there are light operators in the CFT spectrum. By light here, we do not mean light in the sense that the mass of the bulk field must be close to the BF bound. We simply mean that the CFT spectrum should contain an operator $\chi$ with
\be
\frac{5}{2}\pi \sqrt{\frac{c}{3}}>\sqrt{18}\pi (\alpha_\chi + \bar \alpha_\chi )
\ee
Using the definition for $\alpha_\chi$, this translates to

\begin{equation}
    \Delta_\chi < \frac{35 (c-1)}{432} = \frac{c-1}{12}\left(1-\frac{1}{6^2}\right).
\end{equation}
In the presence of matter, we thus seem to expect a new gravitational saddle that would dominate over the genus-two wormhole. We discuss this further in the discussion section.

An interesting remark is that these weights match the conformal dimension of an operator whose bulk interpretation is a conical defect geometry with deficit angle $2\pi(1-1/6)$. These operators have been recently studied in the context of pure three-dimensional gravity as a minimal cure for the non-unitarity of the pure gravity partition function, see \cite{Benjamin:2020mfz} for a detailed discussion. Therefore, whether or not we expect such a solution to contribute in pure gravity could even depend on what we mean by ``pure gravity" in AdS$_3$. If we think pure gravity contains no matter fields and also no conical defects, then the new solution should not contribute. If instead we allow conical defects, the solution could contribute.

\section{Discussion}
\label{sec:discussion}

In this paper, we have used the Virasoro crossing kernel to give asymptotic formulas for OPE coefficients in two-dimensional conformal field theories. We were particularly interested in observables involving more than two OPE coefficients, such as genus-$g$ partition functions with $g>2$. From a statistical point of view, these asymptotic formulas encode the higher moments of the distribution of OPE coefficients, beyond the Gaussian approximation. We gave closed-form expressions for these non-Gaussianities in two channels that have natural liftings up to arbitrarily high genus: the skyline and the comb channel. We then discussed the implication of these results for the calculus of squares of partition functions using the ORH ansatz. We found that in some cases, the quartic moments of OPE coefficients can have comparable or even bigger contributions than the connected Gaussian piece. We conclude with some open questions.

\subsection{All asymptotic formulas at genus three}

At genus three, there are five possible channels for the decomposition of a genus-three surface, leading to five different trivalent topologies. In this paper, we have presented asymptotic formulas for three of the five topologies: the skyline, the comb and the channel \rref{mixedchannel}. The two remaining ones are the necklace and the sunset channels, and we have not been able to extract asymptotic formulas for those. The main difficulty comes from the fact that we were not able to find an appropriate cross-channel where only light data enters. This seems to suggest that the moduli space has the following property: when we take the moduli of these two channels to be small, there does not exist a dual channel where all moduli are large. If this is true, it is puzzling at the level of the asymptotic formulas. It would mean that some asymptotic formulas are not at all universal, i.e. that they depend on the full spectrum of the theory. It would be interesting to understand this better, and we hope to return to this question in the future.

\subsection{Lightest field dependence}

Some of the new asymptotic formulas we have derived involve non-universal data of the CFT and thus depend on more than just the central charge of the theory. In particular, we have seen that some asymptotic formulas depend explicitly on the scaling dimension of the lowest dimension operator of the theory. A similar observation was already made for the asymptotic formula $C_{HHL}$ using modular covariance of torus one-point functions \cite{Kraus:2016nwo}.

It is interesting to observe that the lightest operator of the theory can also affect the partition function in ways that seem novel compared to the torus partition function or even the genus-two partition function in the sunset channel. An example of this is the result for the genus-two partition function in the dumbbell channel \rref{dumbbellpartfunc}. One is perhaps more used to the torus partition function, where the lightest field other than the identity gives a correction to the leading answer that is exponentially suppressed at high temperatures. In particular, one has
\be \label{BTZdisc}
Z_{\textrm{torus}}\approx e^{\frac{c}{12} \frac{4\pi^2}{\beta}}\left(1+e^{-\Delta_\chi \frac{4\pi^2}{\beta}}\right) \,,
\ee
as $\beta\to0$. This should be contrasted with the genus-two dumbbell case which behaves as
\be \label{dubmmbelldisc}
Z_{\textrm{dumbbell}}\approx e^{\frac{c}{4} \frac{\pi^2}{\tilde{\beta}}} e^{\Delta_\chi \frac{2\pi^2}{\tilde{\beta}}} \,.
\ee
where we have denoted the modulus $\tilde{\beta}$ as it is no longer a real temperature. We also have assumed $\Delta_\chi\ll c$. Whether or not the lightest operator gives an exponentially small correction or a prefactor correction directly depends on whether the identity can dominate in the cross channel.

From a gravity standpoint, this is puzzling. A naive guess would be to try and interpret this term as a 1-loop determinant of some bulk field dual to $\chi$. The problem is that the product structure of 1-loop determinants implies that all matter fields should contribute in a uniform way, therefore we would expect the prefactor to take the form of a product over all the light fields. This is clearly not the case, as the second lightest operator gives an exponentially small correction ala \rref{BTZdisc}. It thus appears that it must be coming from something different than a 1-loop determinant. Another possibility is that the lightest operator is somehow important to ``support" the geometry. For example, one could imagine that the 1-loop expectation value of the stress-tensor of the lightest field backreacts on the background, and produces this effect. It would be interesting to explore this further.

\subsection{New gravitational saddles?}

One of the most surprising outcomes of our results is the fact that the comb channel quartic contribution leads to an exponentially large saddle for the square of the genus-two partition function, in the dumbbell channel. This new saddle-point should give a connected contribution to $Z^2$ which dominates over the genus-two wormhole.

It is interesting to note that such a saddle depends on the bulk matter content. For pure gravity with no new operators below $\Delta=\frac{c}{12}$, there is no new saddle-point (it is worth mentioning that such a theory may not lead to a consistent CFT). However, in any top-down scenario such as the one discussed in \cite{Maldacena:2004rf} where the genus-two wormhole is perturbatively stable, one has additional matter fields. 

This seems to indicate that there are new solutions to AdS 3d gravity coupled to matter with two asymptotic regions with negative curvature. The existence of matter seems important to support the wormhole, even though the mass of the field is not important for the existence of the saddle-point (as long as it is light enough). The fact that the value of the mass is unimportant is peculiar, and should be contrasted with other types of AdS$_3$ + matter solutions that exist for higher genus surfaces \cite{Belin:2017nze,Dong:2018esp}. In those cases, the fact that the matter field is light (i.e. close to the BF bound) is crucial. It would be interesting to understand this better, and find the new solutions.

\subsection{Does typicality completely predict the wormhole contribution?}

Finally, we speculate on whether or not typicality is expected to completely fix the generating function $\mathcal{Z}(J)$. The ansatz for our generating function is based on the two following assumptions: first, we assume that the change of basis between an operator and that of the Hamiltonian is given by a random unitary, which we effectively average over. Whether it is a Haar-random unitary or has a different measure is not so important, but the potential for this change of basis is only required to be unitary invariant. Second, we have assumed we can insert only single-trace interactions for $J$.

This second point is more subtle, and we currently do not have a good reasoning for this. Note that there is some ambiguity in how we define the various terms anyway, because we are only defining the coefficients as their leading behaviour in the expansion in $e^S$. When doing Haar-averages, one typically picks up factors of the type
\be
\frac{e^S}{e^S+1} \,,
\ee
which we would have called 1. These come from the Weingarten functions that appear in the Haar averages \cite{Belin:2021ibv}. Note that similar factors appear in \cite{Pollack:2020gfa}, and have also been discussed in \cite{Stanford:2021bhl} in the context of unitarity restoration. The way we have constructed our generating function, we are agnostic about the true potential for the unitaries $U$ and are just matching leading order terms. From this point of view, the subleading corrections could either be correcting the single-trace terms or could be pushed completely into double-trace terms. Naturally, doing Haar averages gives a well-defined answer \cite{Belin:2021ibv}, but the question is to what extent we think Haar-averaging is the right prescription.

This question is physically important. If we believe typicality to be true, and moreover believe no multi-trace terms are required in our generating function, then the ansatz \textit{predicts} what the wormhole contribution should be. On the other hand, if one could accurately calculate all the one-loop determinants around wormhole geometries one should be able to check if single-trace terms are enough, or if the wormhole contribution is in some sense defining the nature of double-trace contributions in the generating function. We leave this for future work.

\section*{Acknowledgements}

We are happy to thank Tarek Anous, Alejandra Castro, Shouvik Datta, Daniel Jafferis, Alex Maloney for fruitful discussions and especially Pranjal Nayak and Julian Sonner for many discussions and collaboration on related projects. We would also like to acknowledge Island Hopping 2020 where many useful conversations on this project took place. JdB and DL are supported by the European Research Council under the European Unions Seventh Framework Programme (FP7/2007-2013), ERC Grant agreement ADG 834878.

\appendix
\section{Results for the fusion kernel}
\label{appendixfusion}
In this appendix, we derive several results for the crossing kernel of Ponsot and Teschner \cite{Ponsot:1999uf,Ponsot:2000mt}. We begin by introducing the explicit form of the fusion kernel and study its asymptotics in different heavy limits. The results we discuss here complement those studied in \cite{Collier:2019weq,Collier:2018exn} where different asymptotic limits and properties are reviewed. Throughout the appendices, $n$ and $m$ are non-negative integers.

The crossing kernel $\bbf$ is defined in two parts: a prefactor and an integral
\begin{equation}
\label{explicitfusion}
    \fker{s}{t}{1}{2}{3}{4} = P_b(P_i;P_s,P_t)P_b(P_i;-P_s,-P_t) \int_{C'}\frac{ds}{i}\prod_{k=1}^4 \frac{S_b(s+U_k)}{S_b(s+V_k)}.
\end{equation}
The prefactor $P_b$ is an expression in terms of $\Gamma_b$ functions
\begin{multline}
	P_b(P_i;P_s,P_t) = 
	\frac{\Gamma_b({\frac{Q}{2}}+i(P_s+P_3-P_4))\Gamma_b({\frac{Q}{2}}+i(P_s-P_3-P_4))}
	{\Gamma_b({\frac{Q}{2}}+i(P_t+P_1-P_4))\Gamma_b({\frac{Q}{2}}+i(P_t-P_1-P_4))}
	\\
	\times \frac{\Gamma_b({\frac{Q}{2}}+i(P_s+P_2-P_1))\Gamma_b({\frac{Q}{2}}+i(P_s+P_1+P_2))}
	{\Gamma_b({\frac{Q}{2}}+i(P_t+P_2-P_3))\Gamma_b({\frac{Q}{2}}+i(P_t+P_2+P_3))}
	\frac{\Gamma_b(Q+2iP_t)}{\Gamma_b(2iP_s)}.
\end{multline}
The arguments of the special function in the integrand are 
\begin{equation}
	\begin{split}
		U_1&=i(P_1-P_4),\\
		U_2&=-i(P_1+P_4), \\
		U_3&= i(P_2+P_3),\\
		U_4&=i(P_2-P_3),
	\end{split}
	\qquad
	\begin{split}
		V_1 &= Q/2+i(-P_s+P_2-P_4),\\
		V_2 &= Q/2+i(P_s+P_2-P_4), \\
		V_3 &= Q/2+iP_t, \\
		V_4 &= Q/2-iP_t.
	\end{split}
\end{equation}
The special function $\Gamma_b(x)$, called the Barnes double gamma function, is a meromorphic function with no zeros and poles at $x = -mb -nb^{-1}$. The double sine function $S_b$ is defined as
\begin{equation}
    S_b(x) = \frac{\Gamma_b(x)}{\Gamma_b(Q-x)},
\end{equation}
and has poles at $x = -mb -nb^{-1}$ and zeros at $x =Q +mb +nb^{-1}$. 

The integrand of the kernel has eight semi-infinite lines of poles. The poles to the left come from the sine functions in the numerator at $s = -U_k -nb-mb^{-1}$ and the poles to the right come from the denominator at $s = Q/2 -V_k +nb+mb^{-1}$. The contour of integration $C'$ runs from $-i\infty$ to $i\infty$ and it is such that it passes in between these two families of poles. In many cases, the left and right lines of poles overlap, when this happens the we must add the corresponding residues to the integral.  

\subsection{Four light external operators}

In this section, we present the general strategy we use to study the fusion kernel in different asymptotic limits. We examine the asymptotic behaviour of this kernel when we have external operators with real fixed momenta $P_i$ in the limit $P_s\rightarrow\infty$. At the end of the computation, we comment on the validity of the result when the momenta of the external operators are in the discrete range.  This discussion is a slight generalization of the results and arguments found in \cite{Collier:2018exn}. 

We start with the prefactor. For now, we take $\alpha_t>0$, making the term $\Gamma_b(Q + 2iP_t) = \Gamma_b(2\alpha_t)$ finite. The prefactor has no other poles at large $P_s$ with fixed $P_i$. There are a few zeros that happen at the poles in the denominator, for instance when $i(P_1-P_4) = -Q/2-iP_t - nb-mb^{-1}$, but these can only happen when both $P_1$ and $P_4$ or $P_2$ and $P_3$ are in the discrete regime. For our case, it is safe to do an asymptotic expansion of the prefactor when $P_s\rightarrow \infty$ without paying too much attention to the constants multiplying the exponential contribution. We find that 
\begin{multline}
        \log \left[P_b(P_i;P_s,P_t)P_b(P_i;-P_s,-P_t)\right]\\= -2 P_s^2 \log 4 + 2\pi \left[\frac{Q}{2} + i(P_2-P_4)\right] P_s + 2\left(\frac{Q^2-1}{4}+\sum_{i=1}^4 P_i^2\right)\log P_s+\clo(1).
\end{multline}
Here, we used the following asymptotic series for $\gb(x)$. This result is valid when $|x|\rightarrow \infty$ for fixed $b$ and $x$ in the right half-plane (see \cite{Collier:2018exn}):
\begin{equation}
    \log \gb(x) = -\frac{1}{2}x^2\log x + \frac{3}{4}x^2+\frac{Q}{2}x\log x-\frac{Q}{2}x - \frac{Q^2+1}{12}\log x + \log \Gamma_0(b) + \clo(x^{-1}),
\end{equation}
where $\Gamma_0(b)$ is a constant. We will also use the asymptotics of $S_b(x)$. The following result is valid as $|x|\rightarrow \infty$ for $x$ in the upper half-plane:
\begin{equation}
    \log S_b(x) = \frac{i\pi}{2}x^2+\frac{i\pi}{2}Q x - \frac{i\pi}{12}(Q^2+1)+\clo(x^{-1}).
\end{equation}
For the integrand, we will use this series and the fact that $\log S_b(x) = - \log S_b(Q-x)$.

Understanding the analytic structure of the integrand in the limit of interest is crucial since the exponential suppressions that matter to us appear -- or do not appear -- in the integral. Before picking a particular contour of integration, we do a change of variables $s = \sigma P_s$ and do an asymptotic expansion of the integrand. There are different regions because $S_b(x)$ has different asymptotic expansions on the upper and lower half-planes. The relevant behaviour of the integrand in each region is  
\begin{equation}
\label{leadingApprox}
    \log \frac{S_b(s+U_k)}{S_b(s+V_k)} \sim
    \begin{cases}
    + 2 i \pi Q \sigma P_s & \Im \sigma>1\\
    -\pi(i \sigma^2+2\sigma)P_s^2 + 2\pi i Q \sigma P_s & 0<\Im \sigma <1\\
    +\pi(i\sigma^2-2\sigma)P^2_s - 2\pi i Q\sigma P_s & -1< \Im \sigma < 0\\
    -2 i \pi Q \sigma P_s & \Im \sigma <-1
    \end{cases}.
\end{equation}

\begin{figure}[t!]
    \centering
    \includegraphics{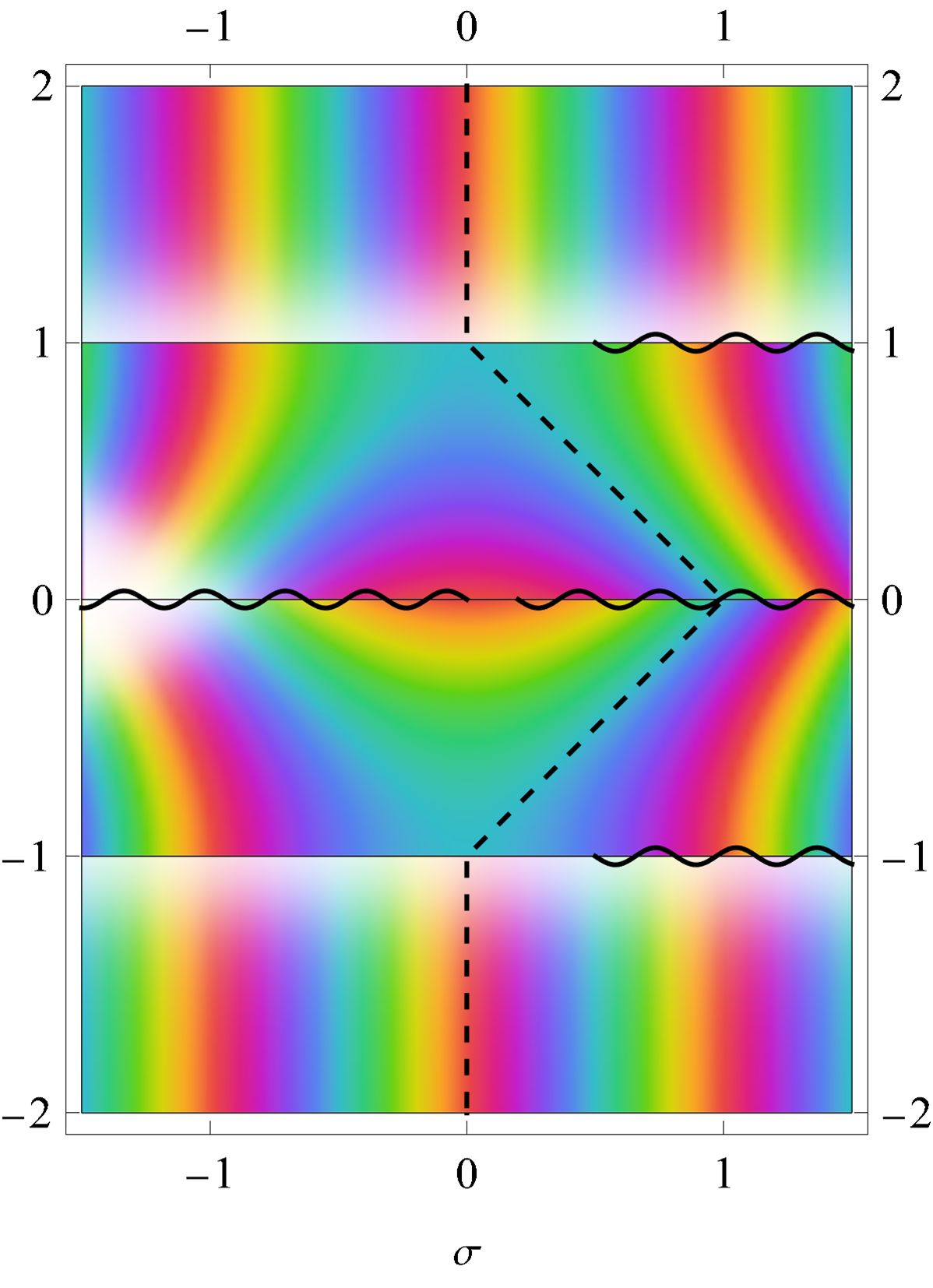}
    \caption{Analytic structure of the integrand in the fusion kernel with $s = \sigma P_s$ and $P_s\rightarrow\infty$. There are eight semi-infinite lines of poles extending to the left and to the right of the integration contour. These families of poles are represented in the plot by black wavy lines. In this example, several of these families overlap so only four wavy lines appear in the image. The dotted line is the integration contour we use to compute the asymptotics of the integral. In the diagram, lighter shading corresponds to a larger absolute value of the integrand in each section.}
    \label{fourpointpoles}
\end{figure}

From the asymptotic behaviour, we see that the integrand is exponentially suppressed in the regions $\Im \sigma>1$ and $\Im \sigma <-1$. So we can take the contour in these two regions to run along the imaginary axis,  see Figure \ref{fourpointpoles}, and it is safe to say that these two parts contribute to a piece of size $\exp{-2\pi Q P_s}$ to the integral.

We first do a naïve approximation for this integral by taking the entire contour to run through the imaginary axis. If the external operators are heavy enough, then no poles cross the axis. To deal with the pole at the origin, we can regularize the integral by doing the integration over the contour $i\mathbb{R}+\epsilon$ and then take the limit $\epsilon\rightarrow 0$ at the end of the computation. With this contour, the integral is exponentially suppressed everywhere because of the factors $\pm 2\pi i Q \sigma P_s$  in the upper and lower half-planes. Thus, we can do a saddle point approximation, evaluate the integrand at the origin, and estimate the integral to be less than a constant times a factor of $\exp{-2\pi i (P_2-P_4)P_s}$. This first result is true, but we can constrain the integral even more. The rapid oscillations coming from the factors $\mp \pi(i \sigma^2 \pm 2\sigma)P_s^2$ in the exponential suggest that the integral should be much less than this first estimation.  

To improve our result, we deform the contour. Note that there are two saddle points at $\sigma = i$ and $\sigma = -i$. They correspond to the saddles of the Gaussians $\exp{\mp \pi(i\sigma^2 \pm 2\sigma) P_s^2}$ dominating the integrand. These saddle points are located to the left of the poles coming from the denominator, see Figure \ref{fourpointpoles}. We can use these saddles to approximate the integral using the method of steepest descent. To achieve this, we deform the contour so that it follows the path $\Re \sigma \pm \Im \sigma =1$. This path is shown in  Figure \ref{fourpointpoles} and corresponds to the path of steepest descent and constant phase. With this new estimate, we conclude that this part of the contour contributes to a term of the size $\exp{-2\pi Q P_s}$. 

The new contour receives corrections from the poles at $s = \alpha_t + n b+ mb^{-1}$ in the real axis. These poles come from term $S_b(s + Q/2 - iP_t) = S_b(s+Q-\alpha_t)$. The important part of the integrand, valuated at $s = \alpha_t + n b+ mb^{-1}$, is given by
\begin{equation}
 -\log S_b(s + V_1)S_b(s+V_2)  \sim -2 \pi \big[s + i(P_2-P_4)\big] P.
\end{equation}
Here, we used the Taylor series of $S_b(x)$ around $x = Q$
\begin{equation}
    S_b(x) = \frac{1}{2\pi}(Q-x),
\end{equation}

From this, we find that the entire integral is dominated by the leftmost pole, with $n=0,m=0$. We have shown that
\begin{equation}
    \log \int_{C'}\frac{ds}{i}\prod_{k=1}^4 \frac{S_b(s+U_k)}{S_b(s+V_k)} \sim    -2 \pi \big[\alpha_t + i(P_2-P_4)\big] P_s.
\end{equation}

If $\alpha_t$ is in the continuum regime, we have to add a second pole to our expressions, coming from $s = Q-\alpha_t+nb+mb^{-1}$. The new pole is the same as this one but with $\alpha_t \rightarrow Q-\alpha_t$ or, $P_t \rightarrow -P_t$. Putting these results together, we find that 
\begin{equation}
\label{asymptoticfourpointkernel}
    \log \fker{s}{t}{1}{2}{3}{4} = -2 P^2_s\log 4 + \pi(Q-2\alpha_t)P_s + 2\left[-\frac{3Q^2+1}{4}+\sum_{i=1}^4 h_i\right]\log P_s + \clo(1).
\end{equation}

This result extends to the case where some or all of the external operators are light. The reason is that the integral is always dominated by the same pole, even if there are fine-tuned parameters.

Explicitly, the exponential suppression of the kernels reads
\begin{equation}
    \log\fker{s}{H}{1}{2}{3}{4}-\log \fker{s}{L}{1}{2}{3}{4} 
    = -2\pi(\alpha_H-\alpha_L)P_s + \clo(1).
\end{equation}
Note that the exponential suppression comes from the integral; the prefactor has no information about $\alpha_t$ at large $P_s$. 

\subsection{All heavy momenta}
\label{allheavymomenta}

\begin{figure}[t!]
    \centering
    \includegraphics{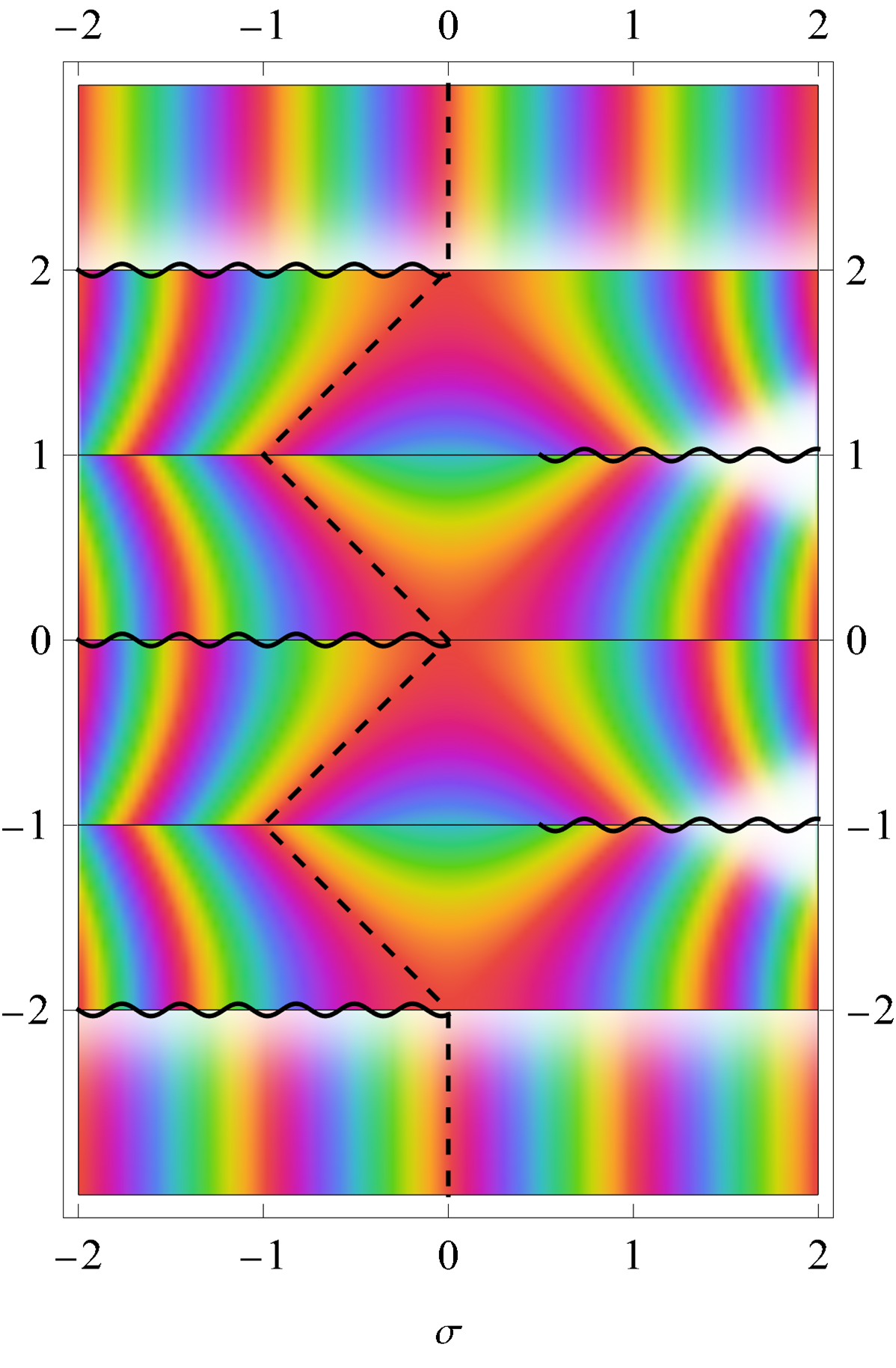}
    \caption{Analytic structure of the integrand in the fusion kernel in the limit where all operators are heavy. The contour of integration is shown with a dotted line. The integral is dominated by the two poles at the origin.}
    \label{polesallheavy}
\end{figure}

We now study the asymptotics of the fusion kernel when all the momenta are heavy. Specifically, we focus on the limit $P_i - \delta_i =  P_s -\delta_s = P_t-\delta_t = P \rightarrow \infty$. The steps to evaluate this kernel are the same as in the previous section. We begin by doing an asymptotic expansion of the prefactor:
\begin{equation}
        \log \left[P_b(P_i;P_s,P_t)P_b(P_i;-P_s,-P_t)\right] = 2\pi Q P +2(\delta_s-\delta_t)P\log\frac{27}{16}+\clo(1).
\end{equation}
This time, the prefactor has no poles or zeros. 
The second step is to evaluate the integral. The relevant terms in the asymptotic expansion of the integrand are
\begin{equation}
    \log \frac{S_b(s+U_k)}{S_b(s+V_k)} \sim
\begin{cases}
    + 2\pi i Q\sigma P & \Im \sigma>2\\
   +\pi (i\sigma^2+4\sigma)P^2+\pi i Q \sigma P -2\pi Q P & 1<\Im \sigma <2\\
   -\pi i \sigma^2 P^2+\pi i Q \sigma P-2\pi Q P& 0<\Im \sigma<1\\
   +\pi i \sigma^2 P^2 - \pi i Q\sigma P -2\pi Q P & -1< \Im \sigma < 0\\
   -\pi (i\sigma^2-4 \sigma)P^2-\pi i Q \sigma P -2\pi  QP & -2 <\Im \sigma <-1\\
   -2\pi i Q\sigma P &\Im\sigma<-2.
\end{cases}.
\end{equation}
The analytic structure of the integrand is summarized in Figure \ref{polesallheavy}. There are three saddle points we can use at our advantage at $\sigma =2i,0$ and $-2i$; the paths extending from $2i$ to $i\infty$ and $-2i$ to $-i\infty$ are exponentially suppressed. The entire integral is of the order of $\exp{-2\pi Q P}$ and it is dominated by the two poles at the origin $s = -i(P_1-P_4)$ and $s = -i(P_2-P_3)$. The important contributions to the integral are
\begin{equation}
    \left(\log \frac{S_b(s+U_2)S_b(s+U_3)}{S_b(s+V_1)S_b(s+V_2)S_b(s+V_3)S_b(s+V_4)} \right)_{s \; = \; -U_1,-U_3}=  -2\pi Q P +\clo(1).
\end{equation}
Both poles have the same asymptotic behaviour. The full kernel then reads
\begin{equation}
\label{allheavy}
    \log \fker{s}{t}{1}{2}{3}{4} = 2(\delta_s-\delta_t)P \log \frac{27}{16} + \clo(1).
\end{equation}

\subsection{Two light external operators}
\begin{figure}[t!]
    \centering
    \includegraphics{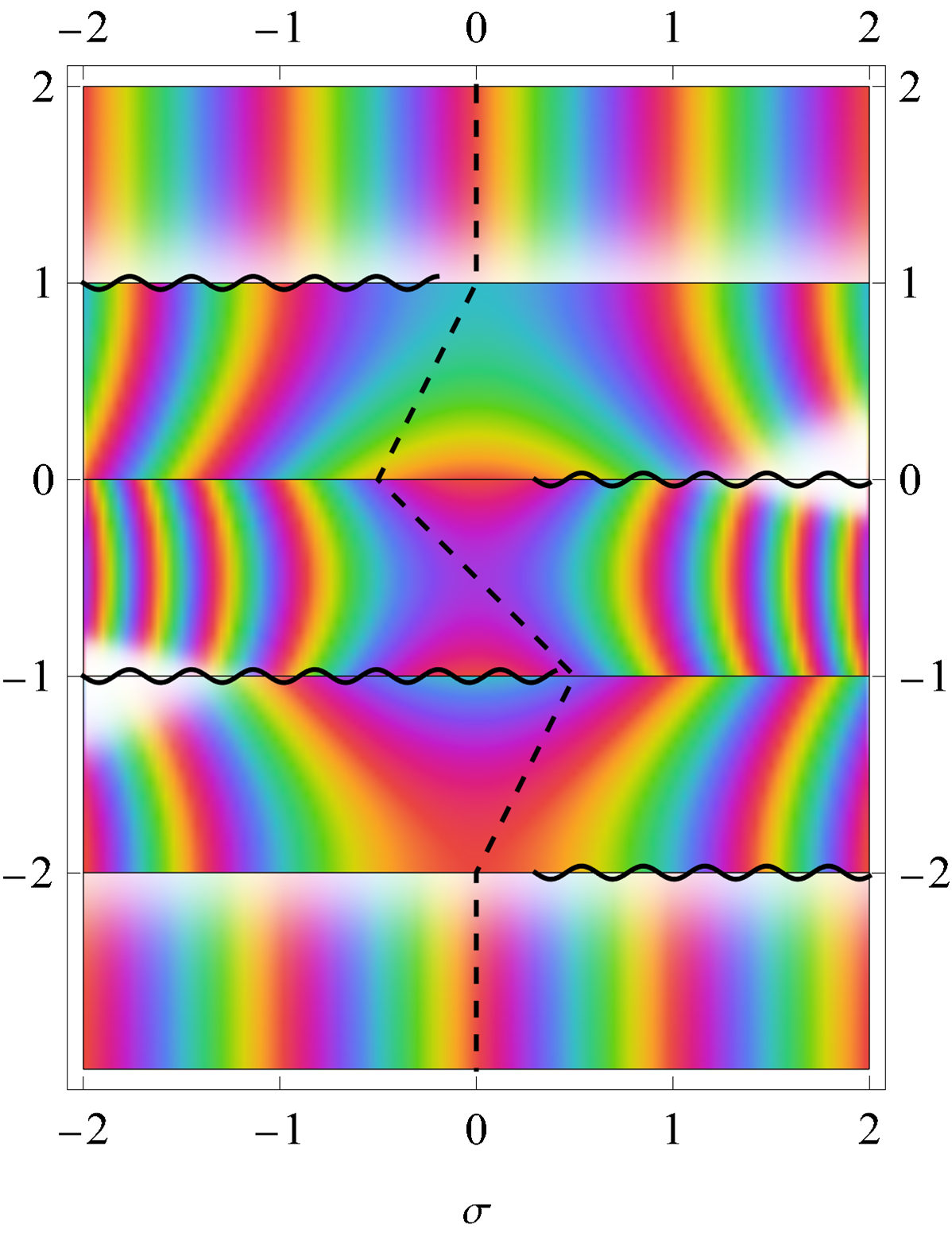}
    \caption{Analytic structure of the integrand in the fusion kernel in the limit $P_s-\delta_s = P_1-\delta_1=P_2-\delta_2 = P\rightarrow\infty$ and light momenta not equal to the identity. This time integral is dominated by its saddle point at $\sigma=-i/2$.}
    \label{twolightextenalanalytic}
\end{figure}

Finally, we examine the asymptotics of the fusion kernel when $P_s-\delta_s = P_1-\delta_1=P_2-\delta_2 = P\rightarrow\infty$ and all light momenta are not equal to the identity. In this limit, the prefactor has no poles and also no zeros. Its asymptotic series is given by 
\begin{multline}
\log \left[P_b(P_i;P_s,P_t)P_b(P_i;-P_s,-P_t)\right] =  \left(-4\log2+\frac{9 \log3}{2}\right)P^2\\+\left(\pi Q +[\delta_1+\delta_2]\log 27-\delta_s \log \frac{256}{27}\right)P +\left(\frac{1+7Q^2}{6}-4h_t\right)\log P +\clo(1).
\end{multline}
The relevant asymptotics of the integrand are 
\begin{equation}
    \log \frac{S_b(s+U_k)}{S_b(s+V_k)} \sim
    \begin{cases}
    +2\pi i Q\sigma P - \pi Q P  &\Im \sigma >1\\
    +\pi(i\sigma^2+2\sigma)P^2+\pi i Q\sigma P-2\pi QP & 0<\Im \sigma <1\\
    -2\pi (i\sigma^2-\sigma)P^2 + \pi i Q \sigma P -2\pi Q P & -1<\Im \sigma <0\\
    +\pi(i\sigma^2-4\sigma)P^2-2\pi i Q\sigma P + \pi Q P & -2< \Im \sigma <-1 \\
    -2\pi i Q \sigma P +\pi Q P & \Im \sigma < -2.
    \end{cases}.
\end{equation}
The analytic structure of the integrand is summarized in Figure \ref{twolightextenalanalytic}. Note that this time our contour is avoiding the poles of the integrand. Using the saddle points at $\sigma = i$ and $\sigma = -2i$, and taking the path of steepest descent, we conclude that the contours going from $0$ to $i \infty$ and $-i$ to $-i\infty$ are of the order of $\exp{-3\pi Q P}$. This integral is well approximated by the saddle point at $\sigma = -i/2$. We find that 
\begin{equation}
    \left(\log  \prod_{k=1}^4 \frac{S_b(s+U_k)}{S_b(s+V_k)}\right)_{s\;=\;-iP/2} =  -\frac{3}{2}\pi Q P +\clo(1).
\end{equation}
Note that a simple saddle point approximation without the deformation of the contour would have given a term of the order of $\exp{-\pi Q P}$. Putting these results together, we find that 
\begin{multline}
    \log \fker{s}{t}{1}{2}{3}{4} = \left(-4 \log 2+\frac{9 \log 3}{2}\right)P^2+ \left(-\frac{\pi Q}{2} +[\delta_1+\delta_2]\log 27-\delta_s \log \frac{256}{27}\right)P\\ +\left(\frac{1+7Q^2}{6}-4h_t\right)\log P +\clo(1).
\end{multline}

\subsection{Delta functions}
\label{deltafunctions}

There are a few limits in which  the fusion kernel behaves as a delta function. In this section, we study two of such cases.

\subsubsection{In the skyline channel}
We start our asymptotic analysis with the kernel
\begin{equation}
    \fker{3}{\bbi}{\bbi+\epsilon}{1}{1}{\bbi +\epsilon} = \rho_0(P_3)C_0\left(P_1,i\left(\frac{Q}{2}-\epsilon\right),P_3\right).
\end{equation}
In the limit $P_1\rightarrow \infty$. The function $C_0(P_1,P_2,P_3)$ is a meromorphic function of its arguments that encodes the behaviour of the average squared OPE heavy coefficients, see \cite{Collier:2019weq} for more details. The analytic structure of this function is determined by its poles at $P_1 = i(Q \pm n b + m b^{-1})/2$, its zeros at $P_1 = P_2+P_3 \pm i Q/2 +i(m b + n b^{-1})$, and its symmetries: invariance under reflections $P_i \rightarrow - P_i$ and permutations of its variables $(P_1,P_2,P_3)$. A closed expression for $C_0$ is 
\begin{equation}
    C_0(P_1,P_2,P_3) = \frac{1}{\sqrt{2}}\frac{\gb(2Q)}{\gb(Q)^3} \frac{\prod_{\pm\pm\pm} \gb(\frac{Q}{2}\pm i P_1 \pm i P_2 \pm i P_3)}{\prod_a \gb(Q+2i P_a)\gb(Q-2iP_a)}.
\end{equation}
In the limit of interest, $C_0(P_1,P_3,\bbi)$ is zero unless $P_1 = P_3$. Here, two poles cancel the zero and the function evaluates to infinity. We can see the delta function contribution by carefully analyzing the limits $\epsilon, (P_1-P_3) \rightarrow 0$ and $P\rightarrow \infty$. We divide $C_0$ into three parts
\begin{align}
\nonumber
&C_0(P_1,\bbi +\epsilon,P_3) \\ \nonumber &= \left(\frac{\gb\big(\epsilon\! +\! i [P_1-P_3]\big)\gb\big(\epsilon\!+\!i[P_3-P_1]\big)}{\gb(2\epsilon)}\right)\left(\frac{\gb(2Q)\gb\big(Q\!-\!\epsilon + i[P_1-P_3]\big)\gb\big(Q\!-\!\epsilon + i[P_3-P_1]\big)}{\sqrt{2}\gb(Q)^3\gb(2Q-2\epsilon)}\right)\\
&\times \left(\frac{\gb\big(Q-\epsilon-i[P_1+P_3]\big)\gb\big(Q-\epsilon +i[P_1+P_3]\big)\gb\big(\epsilon -i[P_1+P_3]\big)\gb\big(\epsilon +i[P_1+P_3]\big)}{\gb\big(Q-2iP_1\big)\gb\big(Q+2iP_1\big)\gb\big(Q-2iP_3\big)\gb\big(Q+2iP_3\big)}\right).
\end{align}
Using the asymptotic expansions for $\gb(x)$ and the Laurent series: 
\begin{align}
    \gb(x) = \frac{\gb(Q)}{2\pi}\frac{1}{x} + \cdots, \quad x\rightarrow 0,
\end{align}
we find that:
\begin{equation}
    C_0(P_1,\bbi,P_3)= \lim_{\epsilon \rightarrow 0}\frac{1}{\pi} \frac{\epsilon}{(P_1-P_3)^2+\epsilon^2}\frac{e^{-2\pi Q P+\clo(1)}}{\sqrt{2}} = \delta(P_1-P_3)\frac{e^{-2\pi Q P+\mathcal{O}(1)}}{\sqrt{2}}.
\end{equation}
Putting all equations together, we find
\begin{equation}
\label{skylinedelta}
    \fker{3}{\bbi}{\bbi}{1}{1}{\bbi} =  \delta(P_1-P_3)\rho_0(P)e^{-2\pi Q P + \clo(1)} = \delta(P_1-P_3)e^{\clo(1)}.
\end{equation}

\subsubsection{In the comb channel}

This sections shows that
\begin{equation}
    \fker{s}{t}{\bbi}{2}{3}{t} = \delta(P_2-P_s).
\end{equation}
This result is a generalization of the previous for cases where $P_t\neq \bbi$. We discuss the two cases separately because they are controlled by different poles pinching the contour of integration. 

When $P_4=P_t$ and $P_1=i(Q/2-\epsilon)$, we have the following contributions coming from the prefactor
\begin{equation}
    P_b(P_i;P_x,P_t)P_b(P_i;-P_s,-P_t)  = \frac{\gb(i (P_2-P_s) + \epsilon)}{\gb(\epsilon)}\times (\textnormal{regular}).
\end{equation}
If $P_2\neq P_s$, the integral in the definition of $\bbf$ remains finite, and the prefactor vanishes so the fusion kernel is zero. We consider $P_2$ close to $P_s$ and $\epsilon$ to be small. In this limit, the poles at $s = -U_1 = Q/2-\epsilon+i P_4$, $s = Q-V_1= Q/2-i(P_2-P_s)+iP_4$ and $s = Q-V_4 = Q/2+iP_t$ pinch the contour of integration. Let us deform the contour to the left to include the pole contribution at $s = Q/2-\epsilon+iP_4$, 
\begin{equation}
    \int_{C'}\frac{ds}{i}\prod_{k=1}^4 \frac{S_b(s+U_k)}{S_b(s+V_k)} = \frac{1}{2\pi}\left(\frac{S_b(Q-2\epsilon)}{S_b(Q-\epsilon)S_b(Q-\epsilon+i(P_2-P_s))}\times (\textnormal{regular})+(\textnormal{regular})\right).
\end{equation}
The regular terms in the sum correspond to the deformed contour integral, these terms will vanish as we take $\epsilon \rightarrow 0$ and only the singular term contributes. Using the Laurent expansions of $\gb$ and $S_b$ and simplifying the regular terms in front of the singular contributions, we find that 
\begin{equation}
\label{combdelta}
    \fker{s}{t}{\bbi}{2}{3}{t} = \lim_{\epsilon \rightarrow 0 } \frac{1}{\pi} \frac{\epsilon}{(P_2-P_s)^2+\epsilon^2}   = \delta(P_2-P_4).
\end{equation}
All terms cancel out and we are left with a delta function in the limit.  

\section{Results for the modular kernel}
\label{appendixmodular}

Here, we derive the asymptotics of the modular $\sker{P_1}{P'}{P_0}$ kernel for heavy operators $P_1 - \delta_1 = P_0 -\delta_0 = P$ and fixed but discrete $P'$. We begin with the definition of the kernel due to Teschner \cite{Teschner:2003at}:
\begin{equation}
\begin{split}
\label{skernel}
	\bbs_{PP'}[P_0] &= \frac{\rho_0(P)}{S_b(\frac{Q}{2}+iP_0)}
	\frac{\Gamma_b(Q+2iP')\Gamma_b(Q-2iP')\Gamma_b(\frac{Q}{2}+i(2P-P_0))\Gamma_b(\frac{Q}{2}-i(2P+P_0))}{\Gamma_b(Q+2iP)\Gamma_b(Q-2iP)\Gamma_b(\frac{Q}{2}+i(2P'-P_0))\Gamma_b(\frac{Q}{2}-i(2P'+P_0))}\\&\eqspace{2cm}\times
	\int_{C'} \frac{d s}{i}e^{-4\pi P's} \frac{S_b\left(s+\frac{Q}{4}+i\left(P+\frac{P_0}{2}\right)\right)S_b\left(s+\frac{Q}{4}-i\left(P-\frac{P_0}{2}\right)\right)}
	{S_b\left(s+\frac{3Q}{4}+i\left(P-\frac{P_0}{2}\right)\right)S_b\left(s+\frac{3Q}{4}-i\left(P+\frac{P_0}{2}\right)\right)}\\
    &= Q_b(P,P',P_0) \int_{C'}\frac{ds}{i}e^{-4\pi P' s} T_b(s,P,P_0).
\end{split}
\end{equation}
Where the contour $C'$ is taken to be the imaginary axis, and $\rho_0(P) = 4 \sqrt{2} \sinh(2\pi b P)\sinh(2\pi b^{-1}P)$. This integral representation only converges when 
\begin{equation}
    \frac{1}{2}\Re(\alpha_0) < \Re(\alpha')<\Re\left(Q-\frac{1}{2}\alpha_0\right). 
\end{equation}
Since our convention is $0<\Re(\alpha)<Q/2$, we only need to worry about the first inequality; the second one is always satisfied. Outside this range, the kernel is defined via  an analytic continuation using the following shift relation
\begin{multline}
\label{shift}
	2\cosh(2\pi b P)\bbs_{PP'}[P_0]=\left(
	\frac{\Gamma(b(Q+2iP'))\Gamma(2ibP')}{\Gamma\left(b\left[\frac{Q}{2}+i(2P'-P_0)\right]\right)\Gamma\left(b\left[\frac{Q}{2}+i(2P'+P_0)\right]\right)}
	\bbs_{P,P'-i\frac{b}{2}}[P_0]\right.\\
	\left. +\frac{\Gamma(b(Q-2iP'))\Gamma(-2ibP')}{\Gamma\left(b\left[\frac{Q}{2}-i(2P'+P_0)\right]\right)\Gamma\left(b\left[\frac{Q}{2}-i(2P'-P_0)\right]\right)}
	\bbs_{P,P'+i\frac{b}{2}}[P_0]\right).
\end{multline}

We start by assuming $\alpha' > Q/4$, so that the integral representation converges. The asymptotics of the prefactor at large $P$ read 
\begin{multline}
\label{Sheavykernel}
    \log Q_b(P_1,P',P_0) = \left(-4\log 2 + \frac{9 \log 3}{2}\right)P^2 +  \left(\pi Q +2\delta_1 \log\frac{27}{16} +  \delta_0 \log 27\right)P\\+\left(\frac{1+7Q^2}{6}-4h'\right)\log P+\clo(1).
\end{multline}
Note that $\alpha'$ does not appear in the prefactor. The estimation of the integrand follows as usual. Doing the change of variables $s =\sigma P$ we find the following relevant asymptotics
\begin{equation}
\log e^{-4\pi P' s} T_b(s,P_1,P_0)  \sim \begin{cases}
    2\pi \sigma P^2 + i\pi (4\alpha' - Q)\sigma P & \Im \sigma > \frac{3}{2}\\
   -\pi \left(i\sigma^2+\sigma \right)P^2 -i \pi \left(\frac{3}{2}Q-4\alpha'\right)\sigma P
    -\frac{3}{4}\pi Q P 
   & \frac{1}{2}< \Im \sigma < \frac{3}{2}\\
      -2\pi i (Q-2\alpha')\sigma P  -\pi QP & -\frac{1}{2}<\Im \sigma < \frac{1}{2}\\
   -\pi \left(i\sigma^2-\sigma\right)P^2 -i \pi \left(\frac{5}{2}Q-4\alpha'\right)\sigma P
    -\frac{3}{4}\pi Q P 
   &-\frac{3}{2}<\Im \sigma < -\frac{1}{2}\\
   -2\pi \sigma P^2 -i\pi (3 Q- 4\alpha')\sigma P & \Im \sigma <- \frac{3}{2}  
\end{cases}.
\end{equation}
This integral is exponentially suppressed above and below $|\Im \sigma|> 3/2$. This time, we have saddle points  at $\sigma =\pm i/2$. The two saddles are related via $\alpha' \rightarrow Q-\alpha'$, since our convention is $\Re(\alpha')<Q/2$, the dominant saddle point is at $\sigma = i/2$ where the integrand is of order $\exp{-2\pi \alpha' P}$. As with the fusion kernel, we follow the contour of constant phase and steepest descent that passes through the two saddle points. The result is that the integral is dominated by its value at $\sigma = i/2$. There are no poles crossing or pinching the contour of integration.

Putting these results together we find that the modular kernel, in this limit, is given by 
\begin{multline}
\label{Sasymptotics}
    \log\sker{P_1}{P'}{P_0} = \left(-4\log 2 + \frac{9 \log 3}{2}\right)P^2 +  \left(\pi (Q-2\alpha') +2\delta_1 \log\frac{27}{16} +  \delta_0 \log 27\right)P\\+\left(\frac{1+7Q^2}{6}-4h'\right)\log P+\clo(1).
\end{multline}
This result holds for $\alpha' > Q/4$. To extend this result for all $\alpha'$ we must use the shift relation. For $\alpha'>Q/4$, the relationship reads:
\begin{equation}
    \sker{P_1}{\alpha'}{P_0} \sim e^{-b\pi P}\left( e^{(1+b Q-4b\alpha')\log P}\sker{P_1}{\alpha'+\frac{b}{2}}{P_0} +e^{(1-3b Q + 4 b\alpha')\log P}\sker{P_1}{\alpha'-\frac{b}{2}}{P_0}\right). 
\end{equation}
Here we have neglected the multiplicative order-one coefficients. We have seen that for $\alpha'> Q/4$, the kernel is exponentially suppressed in $\alpha'$, so we can neglect the contribution from the $\sker{P}{\alpha'+\frac{b}{2}}{P_0}$ kernel. The shift relationship now reads
\begin{equation}
     \log \sker{P_1}{\alpha'-\frac{b}{2}}{P_0}\sim 
     b\pi P-(1-3b Q + 4 b\alpha')\log P +\log \sker{P_1}{\alpha'}{P_0}.
\end{equation}
It is a nontrivial check on the asymptotics of the heavy kernel to verify that this formula works for $\alpha'$ above $Q/4$. The shift relationship also implies that equation \rref{Sasymptotics} is valid for all values of $\alpha'$.

\bibliographystyle{ytphys}
\bibliography{ref}

\end{document}